\documentclass[aps,pre,preprint,superscriptaddress,nofootinbib,floatfix]{revtex4-2}
\usepackage[T1]{fontenc}
\usepackage{lmodern}
\usepackage[hidelinks]{hyperref}
\usepackage{orcidlink} % optional (comment out if not installed)
\usepackage{amsfonts,amsmath,amssymb}
\usepackage{subcaption}
\usepackage{siunitx,physics}
\usepackage{url}
\usepackage{graphicx}
\begin{document}
\title{Beyond directions: Symmetry-aware rotation sets for triaxial diffusion encoding by geometric filter optimization}%Beyond directions: Rotation sets for triaxial diffusion encoding by geometric filter optimization (GFO)}

\author{Sune N{\o}rh{\o}j Jespersen\orcidlink{0000-0003-3146-4329}}
\email{sune@cfin.au.dk}
\affiliation{Department of Clinical Medicine, Aarhus University, Denmark}
\affiliation{Department of Physics and Astronomy, Aarhus University, Denmark}
\thanks{Present address: CFIN, Universitetsbyen 3, Building 1710, 8000 Aarhus, Denmark.}
\author{Filip Szczepankiewicz\orcidlink{0000-0002-5251-587X}}
\affiliation{Department of Medical Radiation Physics, Lund University, Sweden}
\begin{abstract}\noindent\textbf{Purpose:}
To improve the accuracy of  diffusion-weighted powder average signals for diffusion encoding with arbitrary b-tensors. \\
\textbf{Methods}
We identify an intrinsic dihedral ($D_2$) symmetry of diffusion signals for arbitrary diffusion encoding, which defines their natural signal space (a quotient of 3D rotations). Based on this, we propose a method to generate optimal rotation sets that are applied to the diffusion-encoding gradient waveform to yield powder averages with maximal accuracy. The method, termed ``Geometric Filter Optimization'' (GFO), amounts to designing a sampling filter that is approximately flat over the relevant part of the associated frequency space. We characterize the filter properties and benchmark performance in terms of the accuracy and precision of powder averages and higher-order rotational invariants, including comparison with spherical designs and electrostatic-repulsion-based designs defined on the same space.\\
\textbf{Results}
 We found that GFO leads to marked improvements in precision and accuracy in powder averaging over diffusion encoding b-tensors, including axisymmetric and triaxial configurations. For higher-order rotational invariants, the performance was more nuanced, with GFO, electrostatic repulsion, and spherical designs exhibiting different trade-offs in bias and precision depending on $b$ and $N$.
\\
\textbf{Conclusion}
A fundamental $D_2$-symmetry of tensor-valued diffusion encoding was shown to constrain its rotational structure and guide the design of optimal rotation sets. This yielded GFO, which provides an efficient recipe for obtaining orientations for powder averaging of signals with axisymmetric and triaxial diffusion encoding. It places no additional demands on gradient system performance and can be used to shorten scan time.
\end{abstract}
\maketitle
\footnotetext{\textbf{Abbreviations: }{DDE, Double Diffusion Encoding; LTE, Linear b-Tensor Encoding, STE, Spherical b-Tensor Encoding, SM, Standard Model; SMEX, Standard Model with EXchange; NEXI, Neurite Exchange Imaging; fODF, fiber Orientation Distribution Function; ESR, electrostatic repulsion; GFO, geometric filter optimization}}
\section{Introduction}
The encoding in diffusion MRI (dMRI) is most commonly performed along a single direction per shot; the field gradient on each axis is applied in a synchronous manner to sensitize the signal to motion along that direction \cite{stejskal1965}. Although this conventional approach has had tremendous success, it is unable to probe certain features of tissue microstructure, which has motivated the development of alternatives.

Tensor-valued diffusion encoding, sometimes referred to as multidimensional dMRI, employs multiple pulsed field gradients (multiple diffusion encoding \cite{Cory99,shemeshConventionsNomenclatureDouble2016}) or continuously modulated asynchronous gradient waveforms, to encode diffusion along more than one direction per shot \cite{szczepankiewiczGradientWaveformDesign2021}. Unlike conventional diffusion encoding, this encoding can no longer be described by a single vector but rather requires a second-order tensor that can have a rank of up to three. Strictly, while this b-tensor description is complete for Gaussian diffusion compartments,  non-Gaussian also depends on additional properties of the diffusion-encoding waveform. For b-tensors with rank larger than one, the diffusion encoding, in addition to its strength and direction, also has a ``shape'' \cite{topgaardMultidimensionalDiffusionMRI2017,westinQspaceTrajectoryImaging2016}. Indeed, it is the combination of multiple b-tensors of different shapes that enables the measurement of microscopic fractional anisotropy \cite{lasicMicroanisotropyImagingQuantification2014,jespersenOrientationallyInvariantMetrics2013} and the disentanglement of isotropic from anisotropic diffusional kurtosis \cite{szczepankiewiczLinkDiffusionMRI2016} and microscopic kurtosis\cite{henriquesCorrelationTensorMagnetic2020}. Acquiring data with multiple b-shapes has also been shown to lift the degeneracy in the Standard Model parameter estimation \cite{coelhoResolvingDegeneracyDiffusion2019,reisertUniqueAnalyticalSolution2019,coelhoReproducibilityStandardModel2022,lampinenProbingBrainTissue2023}.

Several methods in dMRI are based on the analysis of the so-called signal powder average \cite{bakREPULSIONNovelApproach1997,jespersenOrientationallyInvariantMetrics2013,lasicMicroanisotropyImagingQuantification2014,kadenQuantitativeMappingPeraxon2016,novikovRotationallyinvariantMappingScalar2018,szczepankiewiczTensorvaluedDiffusionEncoding2019}. Briefly, the powder-average is the directional average of the signal on a given b-shell, intended to approximate the signal in a substrate in which the orientations of all domains are isotropically distributed, like in a powder \cite{edenComputerSimulationsSolidstate2003}. In turn, this enables a compact mathematical representation of the signal which is generally easier to use in estimation \cite{mckinnonDependenceBvalueDirectionaveraged2017,veraartNonivasiveQuantificationAxon2020}. Beyond the powder average, appropriately rotated acquisitions can also be used to estimate higher-order rotational invariants of the signal\cite{novikovRotationallyinvariantMappingScalar2018}, which provide additional information about microstructure and likewise require accurate rotational sampling. When using encoding b-tensors that are axisymmetric, e.g., conventional diffusion encoding or tensor-valued encoding with only one or two unique eigenvalues, the set of rotations that yield an accurate powder average is straightforward to produce using, for example, electrostatic repulsion of point charges on the mantle of a sphere \cite{bakREPULSIONNovelApproach1997,jonesOptimalStrategiesMeasuring1999}. By contrast, for \textit{triaxial encoding}---b-tensors with three distinct eigenvalues without a symmetry axis (Fig. ~\ref{fig:B})---an isotropic distribution of encoding orientations requires additional considerations\cite{grafSamplingSetsQuadrature2009,jespersenIsotropicSamplingTensorencoded2025,jespersenOrientationallyInvariantMetrics2013,westinPlatonics2020}. We expect such  rotation sets to be relevant for example for recent work on rotational invariants under arbitrary b-tensor encoding \cite{coelhoGeometryCumulantSeries2026} and for generalized Standard Model frameworks, in which the signal is represented on the rotation group rather than the sphere \cite{CoelhoISMRM2026}.

The aim of this work is to identify the relevant symmetry structure of diffusion signals under diffusion encoding with general sets of eigenvalues, and to use it to optimize sets of rotations for powder averaging and higher-order rotational invariants.

\section{Theory}\label{theory}
Here we outline the theoretical basis of the method, while presenting some of the lengthier derivations to the Appendix.
\subsection{Harmonic analysis on \texorpdfstring{$SO(3)$}{SO(3)}}
We begin by briefly listing some basic but necessary properties of harmonic analysis on the rotation group $SO(3)$, and note the factorization of kernel and fiber orientation distribution functions (fODF) in the Standard Model\cite{novikovQuantifyingBrainMicrostructure2019} extended to tensor-valued diffusion encoding and triaxial diffusion tensors in the Appendix. See also recent work by Coelho et al \cite{coelhoGeometryCumulantSeries2026,CoelhoISMRM2026} for group theoretical analysis of the diffusion signal, including explicit calculation of SO(3) Fourier components and  rotational invariants.

In this work, we use $h$ and $g$ to denote both the abstract elements of $SO(3)$ and, by a slight abuse of notation, their corresponding 3$\times$3 rotation matrices in the defining representation.  Integrals over $SO(3)$ are normalized such that
\begin{equation*}
     \int_{SO(3)}\!\! \mathrm{d}h=1.
\end{equation*}
For example, we can take $\mathrm{d}h = \sin\beta\,\mathrm{d}\alpha\,\mathrm{d}\beta\,\mathrm{d}\gamma/(8\pi^2)$, known as the Haar-measure, where $\alpha,\beta,\gamma$ are ZYZ active rotation Euler-angles $R(\alpha,\beta,\gamma) = R_z(\alpha)R_y(\beta)R_z(\gamma)$.

For functions $f(h)$ from $SO(3)$ to $\mathbb{C}$, we define the Fourier Transform
\begin{align}
\label{eq:FT1}
    \begin{split}f (h) &= \sum_{l=0}^\infty \sum_{m,n=-l}^l (2l+1)\mathcal{D}^{l*}_{mn}(h)f^l_{mn} \\
    & =\sum_l (2l+1)\Tr \left(f^l\mathcal{D}^l(h^{-1})\right)\end{split}\\
    \Updownarrow & \nonumber\\
    \label{eq:FT2}
    & f^l_{mn} = \int_{SO(3)}\!\!  \mathcal{D}^l_{mn}(h)f(h)\,\mathrm{d}h,
\end{align}
where $\mathcal{D}^l(h)$ are the  Wigner matrices, constituting irreducible representations of $SO(3)$, and thereby a complete set by the Peter-Weyl theorem. Throughout, we will make extensive use of their defining property as a unitary representation, $\mathcal{D}^l(h_1h_2)=\mathcal{D}^l(h_1)\mathcal{D}^l(h_2)$, and $\mathcal{D}^l(h_1^{-1}) = \mathcal{D}^l(h_1)^\dagger$ for any $h_1,h_2\in SO(3)$, and their orthogonality
\begin{equation*}
    \int_{SO(3)} \mathcal{D}^{l_1}_{m_1n_1}(h)\mathcal{D}^{l_2*}_{m_2n_2}(h)\,\mathrm{d}h =\frac{\delta_{l_1,l_2}\delta_{m_1,m_2}\delta_{n_1,n_2}}{2l_1+1}.
\end{equation*} 
Note that $f^l$ and $\mathcal{D}^l$ are both complex matrices of size (2l+1)-by-(2l+1). As usual, convolution factorizes for Fourier transforms
\begin{eqnarray}
\label{eq:conv}
    &&F(g)\equiv (f_1\otimes f_2) (g) \equiv \int_{SO(3)}\!\!  f_1(gh^{-1})f_2(h)\,\mathrm{d}h\\ 
    \Updownarrow &&\nonumber \\
    && F^l = f_1^lf_2^l.
\end{eqnarray}
With respect to these Fourier transforms, different conventions concerning the complex conjugation of Wigner matrices and the factor of $2l+1$ are possible --- see e.g. \cite{coelhoGeometryCumulantSeries2026,CoelhoISMRM2026} for another choice. However, as long as one is consistent, this does not affect the results.

\subsection{Powder averaging on \texorpdfstring{$\mathbb{S}^2$}{S2}}
For concreteness, we will exemplify the diffusion-weighted signal assuming anisotropic Gaussian diffusion
\begin{eqnarray}
\label{eq:Gauss}
    S(\mathrm{B}) = \exp(-B_{ij}D_{ij}) = \exp(-\mathrm{B}:\mathrm{D}),
\end{eqnarray}
where we employed the Einstein summation convention for summing over repeated indices. In Eq.~(\ref{eq:Gauss}), D is the diffusion tensor and B is the diffusion weighting tensor (b-tensor), defined in terms of the effective diffusion encoding gradient waveform $\mathbf{G}(t)$ as
\begin{equation}
    B_{ij} = \gamma^2\int_0^T\mathrm{d}t\int_0^t \mathrm{d}t_1\int_0^t \mathrm{d}t_2\, G_i(t_1)G_j(t_2),
\end{equation}
and a b-value defined by its trace, $b= B_{ii}$. However, we note that the theory described herein generalizes to any other diffusion signal model or representation with any functional dependence on the gradients through B, e.g., multi-Gaussian diffusion. 

For linear b-tensor encoding (LTE) with $\text{B}=b\hat{\mathbf{g}}\hat{\mathbf{g}}^\text{T}$, the powder average $(\bar{S})$ of the diffusion-weighted signal 
$(S)$ is an average over the 2-dimensional sphere in 3 dimensions, $\mathbb{S}^2$
\begin{equation}
\bar{S}(b)  
 \equiv  \int_{\mathbb{S}^2}  S(b,\hat{\mathbf{g}})\mathrm{d}\hat{\mathbf{g}},\;\;\;\text{with}\;  \int_{\mathbb{S}^2}  \mathrm{d}\hat{\mathbf{g}}=1.
\end{equation}
In practice, the integral is approximated by a weighted average of signals acquired along some carefully chosen directions, e.g., using the principle of electrostatic repulsion\cite{jonesOptimalStrategiesMeasuring1999,bakREPULSIONNovelApproach1997}, $\bar{S}\approx \sum_iw_iS(\hat{\mathbf{g}}_i)$ with weights $\sum_iw_i=1$. 
In contrast to the true powder average, the estimate will depend on the orientation of the sample. 
\subsection{Powder averaging on \texorpdfstring{$SO(3)$}{SO(3)}: Geometric Filter Optimization}
\begin{figure}[tbp]
  \centering
  \includegraphics[width=\linewidth]{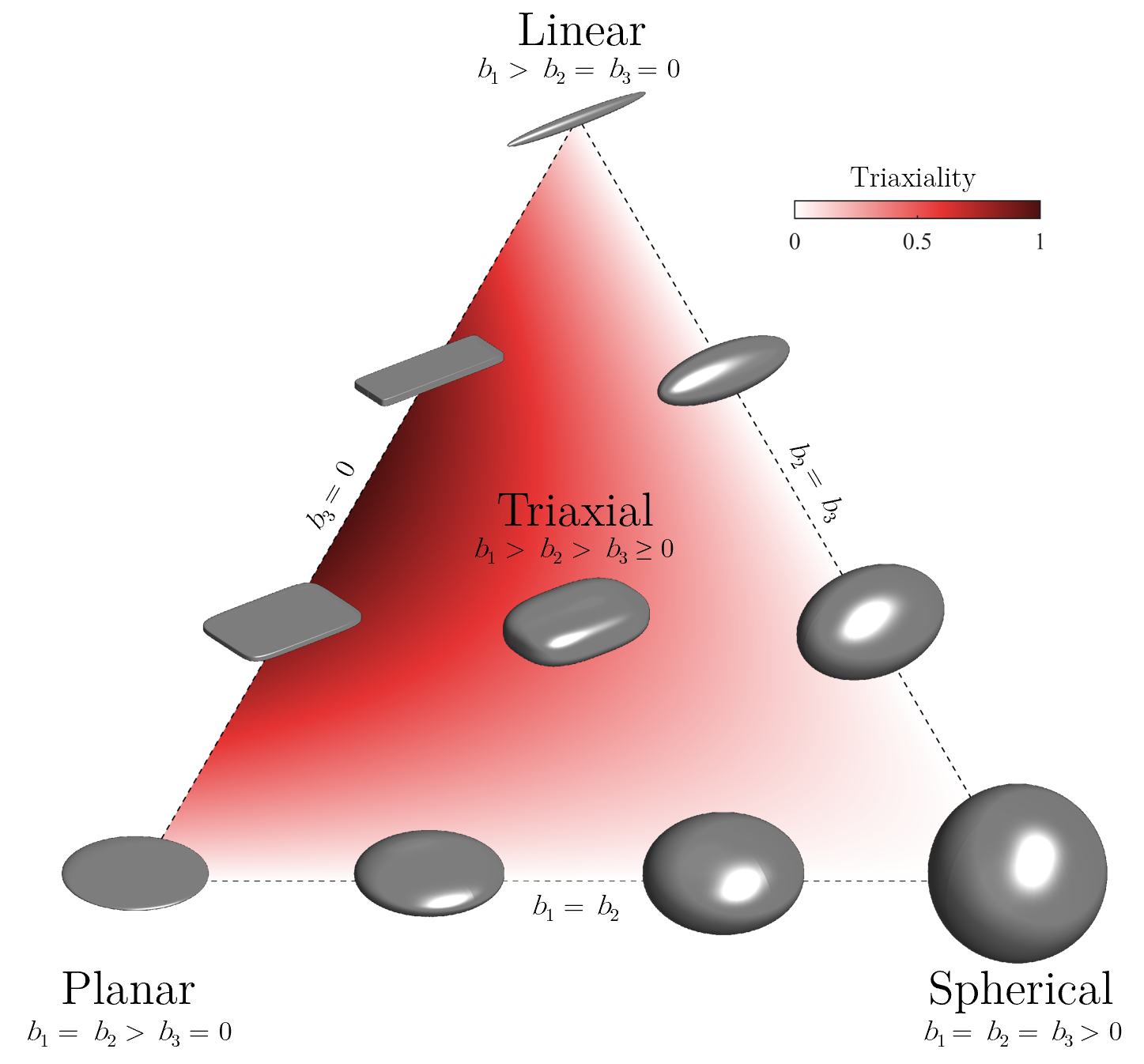}
  \caption{The landscape of b-tensors includes all possible combinations of eigenvalues and can be visualized as glyphs in a triangle wherein the corners and two edges contain the axisymmetric configurations. The remainder of this triangle (red area) contains all triaxial b-tensor shapes, each with three unique eigenvalues. The colormap encodes {\it triaxiality}, a measure of how triaxial a b-tensor is, as defined here by $8(b_1-b_2)(b_2-b_3)/b^2$.}\label{fig:B}
\end{figure}

The natural generalization to arbitrary b-tensor encoding involves an average over all orientations of B, i.e., an average over the rotation group $SO(3)$\cite{jespersenOrientationallyInvariantMetrics2013,jespersenIsotropicSamplingTensorencoded2025}. For that, we define $\mathbb{B}$ as the trace-normalized b-tensor in its principal axis system, i.e., 
\begin{equation}
\mathbb{B} = \frac1{b}\begin{bmatrix}
b_1 & 0 & 0  \\
0 & b_2 & 0\\
0 & 0 & b_3
\end{bmatrix}.
\end{equation}
where the $b$-value is $b=b_1+b_2+b_3$. Note that in terms of normalized linear, planar, and spherical components\cite{topgaardMultidimensionalDiffusionMRI2017}, $b_L/b$, $b_P/b$ and $b_S/b$, we have
\[
\frac{b_S}{b}=\frac{3 b_3}{b}, \quad \frac{b_P}{b}=\frac{2\left(b_2-b_3\right)}{b}, \quad \frac{b_L}{b}=\frac{b_1-b_2}{b}.
\]
The full landscape of b-tensor shapes is illustrated in Fig.~\ref{fig:B}, and we will focus on triaxial b-tensors defined by $b_1> b_2> b_3 \geq 0$. We then consider the signal acquired with $\mathrm{B} = g\mathbb{B}g^{-1}$ as a function of a rotation, $ g\in SO(3)$ 
\begin{equation*}
g\mapsto S(g)\equiv S(b,g\mathbb{B}g^{-1}) \in \mathbb{R},
\end{equation*}
with the dependence on $b$ and $\mathbb{B}$ being implicit.
For any $\mathbb{B}$, the true signal powder average is
\begin{equation}
\label{eq:def_Sbar}
\bar{S}(b,\mathbb{B}) \equiv \int_{SO(3)}\!\!S(g)\, \mathrm{d}g.
\end{equation}
The practical version, which involves a weighted signal average, can be expressed as a convolution with a filter $f(h) = \sum_iw_i\delta(hh_i^{-1})$, where $\delta(hh_i^{-1})$ is the delta function centered at $h=h_i$ on $SO(3)$
 \begin{equation}
 \label{eq:Sbar_discrete}
\bar{S} \approx \widehat{S}(g)\equiv\sum_{i=1}^N w_i S(gh_i^{-1})
=\! \int_{SO(3)}\!\!S(gh^{-1}) f(h)\,\mathrm{d}h,
\end{equation}
and, hence per Eq.~(\ref{eq:conv}), $\widehat{S}^l = S^lf^l$. Thus, the filter is defined in terms of its weights $w_i$ and sampling points $h_i$. As with LTE, while the true orientational average is rotationally invariant, the estimate $\widehat{S}(g)$ above may still depend on the relative orientation of B with respect to the sample, which we have made explicit with the dependence on $g$. Comparing Eqs.~(\ref{eq:def_Sbar}) and \ref{eq:Sbar_discrete} we see that perfect powder averaging is achieved by $f(g)=f_0(g)\equiv1$, or in terms of its $SO(3)$ Fourier-coefficients, $(f_0)^l_{mn} =\delta_{l0}\delta_{m0}\delta_{n0}$. Hence, a good filter is characterized by $f\approx f_0$.  From the filter definition, we can read off its Fourier components directly as $ f^l_{mn}=\sum_iw_i\mathcal{D}^l_{mn}(h_i)$, or expressed in matrix form as $ \mathbf{f} =\mathrm{A}\mathbf{w} $,  where $\mathrm{A}$ contains the Wigner-D matrices $\mathcal{D}^l_{mn}(h_i)$ with $lmn$ across rows (up to some $L$), $i$ across columns, and $\mathbf{f}$ stacks the  $f^l_{mn}$. In terms of dimensions, $\mathbf{w}$ is $N\times 1$, $\mathrm{A}$ is $N_L\times N$ and  $\mathbf{f}$ is $N_L\times 1$, where $N_L = \sum_{l=0}^L(2l+1)^2 =1/3 (1 + L) (1+2L)(3 + 2 L)$. We thus seek to approximate $f$ to the ideal filter $f_0$ by minimizing
\begin{equation}
    \label{eq:opt}
    \lVert \mathrm{V}(\mathbf{f}-\mathbf{f_0})\rVert ^2 = \lVert \mathrm{V}\mathrm{A}\mathbf{w} - \mathrm{V}\mathbf{f_0}\rVert ^2,
\end{equation}
with $\mathrm{V}$ an $N_L\times N_L$ diagonal matrix of user-defined weights across $l$. 
An equivalent formulation for $\mathbb{S}^2$ (LTE) was used in \cite{knutssonAdvancedFilterDesign1999,afzaliComputingOrientationalaverageDiffusionweighted2021,szczepankiewiczMeasurementWeightingScheme2017}, to optimize weights given directions based on electrostatic repulsion. That scheme can be extended to $SO(3)$\cite{jespersenIsotropicSamplingTensorencoded2025}. 

Here, we propose "Geometric Filter Optimization" (GFO), where we instead optimize the \textit{rotations} $h_i$ with\textit{ fixed equal weights} given by $w_i = 1/N$. Then the cost function in Eq.~(\ref{eq:opt}) reduces to (see Appendix)
\begin{multline}
    \label{eq:opt2}
J(\{h_i\}_{i=1\ldots N})=(\mathrm{A}\mathbf{w}-\mathbf{f}_0)^\dagger \mathrm{V}^\dagger \mathrm{V} (\mathrm{A}\mathbf{w}-\mathbf{f}_0)\\
= c + \frac1{N^2}\sum_{j,k=1}^N\sum_{l=0}^L V_{ll}^2 \sum_{m,n=-l}^l \mathcal{D}^l_{mn}(h_j) \mathcal{D}^l_{mn}(h_k)^* \\
= c + \frac1{N^2}\sum_l V_{ll}^2  \sum_{jk} \chi^l\big(h_jh_k^{-1}\big)\\
= c + \frac1{N^2}\sum_l V_{ll}^2  \sum_{jk} \frac{\sin[(2l+1)\beta_{jk}/2]}{\sin(\beta_{jk}/2)},
\end{multline}
\noindent where $c$ is a constant, $\chi^l(h)=\Tr (\mathcal{D}^l)$ are the $SO(3)$ group characters, and $\beta_{jk}$ is the rotation angle of $h_jh_k^{-1}$.  In the quaternion representation, $h \leftrightarrow q$, this simplifies to $\chi^l(h_jh_k^{-1}) = U_{2l}(q_j \cdot q_k)$, where $U_{2l}$ is a Chebyshev polynomial of the second kind. Therefore, in practice, the unit quaternions, with $q$ and $-q$ identified, are the most convenient optimization variables.

To guide the selection of V, we consider the variance of the estimated signal powder average, $\mathrm{Var}[\widehat{S}(g)]$, over the orientations $g$ of B
\[
\mathrm{Var}[\widehat{S}(g)] = \int_{SO(3)}(\widehat{S}(g))^2\,\mathrm{d}g  -  \Big(\!\int_{SO(3)}\!\widehat{S}(g)\,\mathrm{d}g\Big)^2,
\]
which we show in the Appendix to be exactly
\begin{equation}
\mathrm{Var}[\widehat S(g)]
= \sum_{l>0}(2l+1)\sum_{m,n}
\Big|\sum_{n'} S^l_{mn'}\,f^{l}_{n'n}\Big|^2.
\label{eq:SO3-var-exact}
\end{equation}
Defining the rotational invariants $S_l$ (and band powers $E_l$), in analogy to linear b-tensor encoding in the Standard Model \cite{novikovRotationallyinvariantMappingScalar2018}, we get (see Appendix)
\begin{eqnarray}
 E_l(S) &\equiv& \sum_{m,n}|S^l_{mn}|^2 =(2l+1)S_l^2,
\\
 E_l(f) &\equiv& \sum_{m,n}|f^l_{mn}|^2=(2l+1)f_l^2,
\end{eqnarray}
and applying the Cauchy--Schwarz inequality to the inner sum in \eqref{eq:SO3-var-exact},
we arrive at
\begin{equation}
\label{eq:bound}
\mathrm{Var}[\widehat S(g)]
\;\le\; \sum_{l>0}(2l+1)\,E_l(S)\,E_l(f).
\end{equation}
This upper bound shows that the contribution of band $l$ is governed by the product of the band powers of $S$ and of the sampling filter $f$.
On the other hand, up to a constant, our GFO cost, Eq.~(\ref{eq:opt2}), is $\sum_l V_{ll}^2 f_l^2$. Identifying these structures suggests choosing
 \[
 V_{ll}^2 \propto (2l+1)E_l(S),
 \]
with $E_l(S)$  estimated from a representative ensemble of signals. With this choice, minimizing the GFO cost Eq.~(\ref{eq:opt}) directly targets the bands that have the largest contribution to the variance of the estimated powder average, and is therefore expected to simultaneously reduce the variance. Note that the choice of V therefore reflects prior assumptions about the signal ensemble of interest, and can in principle be adapted when more specific prior information about the signal ensemble is available.

A rough generic estimate of the variance in terms of signal power alone can be made by assuming that the orientations are uniform. Then, for $l>0$, $\langle \mathcal{D}^l_{mn}(h_j) \mathcal{D}^{l*}_{mn}(h_k)\rangle = \delta_{jk}/(2l+1)$, and therefore $\langle f_l^2\rangle = (2l+1)/N$. Hence, we can approximate an upper bound for the variance of powder averages, according to
\begin{equation}
    \mathrm{Var}[\widehat S(g)]
\; \lesssim \; \frac1{N}\sum_{l>0}(2l+1)^3\,E_l(S).
\end{equation}
\subsection{A fundamental symmetry of diffusion signals}
So far, we have not exploited any properties of the class of signals we are considering. However,  symmetry properties restrict the Fourier components of the signal, and therefore we can focus our efforts on controlling only the relevant filter components. Diffusion signals that depend only on $g$ via $\mathrm{B}=g\mathbb{B}g^{-1}$ are  invariant to transformations $g\to gK$ for $K\in \{ I,  R_x(\pi),  R_y(\pi), R_z(\pi)\}$, the dihedral group $D_2\subset SO(3)$, since
\begin{equation*}
    (gK)\mathbb{B}(gK)^{-1} = g(K\mathbb{B}K^{-1})g^{-1} = g\mathbb{B}g^{-1},
\end{equation*}
as $K\mathbb{B}K^{-1}$ just inverts the sign of two eigenvectors. The symmetry is thus a simple consequence of the b-tensor glyphs being invariant to $\pi$-rotations around its principal axes. In short, $S(g)$ is right-invariant to $D_2$. It follows that $S^l = S^l\mathcal{D}^l(K)$ for all $K\in D_2$, and that $S^l$ therefore lies in the image of the projector
\begin{equation*}
    P^l = \frac14\sum_{K\in D_2} \mathcal{D}^l(K),\;\;\;(P^l)^2 = P^l,
\end{equation*}
which means that $S^l = S^lP^l$. In the Appendix, we derive these properties and also show that they imply that the entire  $l=1$ sector as well as all $S^l_{mn}$ with odd $n$ vanish. By contrast, and unlike for LTE, the odd-$l$ coefficients do not vanish in general; however, the associated band power is relatively small, which seems plausible because the projector $P^l$ has a lower-dimensional image, and thus fewer modes contribute. Further, while the odd bands are unlocked by triaxiality they remain strongly suppressed for the Gaussian tensor signal class considered here.  Perturbatively, near the axisymmetric edges one finds $E_l=\mathcal{O}(\epsilon^2)$ for odd $l\neq 1$, where $\epsilon=(3/b)\operatorname{min (b_1-b_2,b_2-b_3)}\in[0,1]$ is a small triaxiality parameter measuring the normalized distance to the nearest axisymmetric edge, c.f. Fig.~\ref{fig:B}. Numerically, the suppression is often much stronger across the full shape space, suggesting an additional structural bias of these signals toward even $l$, beyond the edge perturbation argument alone. 

Since the true signal is right-invariant under the action of $D_2$, i.e., $S(gK)=S(g)$ for all $K\in D_2$, it is natural to require the estimated powder average to respect the same invariance. This is achieved by symmetrizing the sampling filter over the right action of $D_2$, defining
\[
f_{D_2}(h) \equiv \frac{1}{4}\sum_{K\in D_2} f(hK).
\]
The corresponding estimate,
\[
\widehat S(g) = \int_{SO(3)} S(gh^{-1})\, f_{D_2}(h)\,\mathrm{d}h,
\]
is then right-invariant by construction. In Fourier space, this symmetrization amounts to projecting the filter coefficients onto the image of the projector so that $f_{D_2}^l = f^l P^l$. Consequently, only the projected components $f^lP^l$ contribute to the estimate. In terms of $SO(3)$, this means we optimize the point wise average of $J$ over $D_2$, such that
\begin{multline}
    \label{eq:opt3}
\frac1{4^N}\sum_{K_1\ldots K_N \in D_2}J(h_1K_1,h_2K_2,\ldots ,h_NK_N)=\\
 c + \frac1{4N^2}\sum_{K\in D_2}\sum_l V_{ll}^2  \sum_{jk} \chi^l\big(h_jKh_k^{-1}\big).
\end{multline}
 Note that this cost function is a function on the quotient space\footnote{Strictly speaking, $N$ copies of the quotient space} $SO(3)/D_2$ whose elements are left-cosets, $gD_2$. The rotations $h_i$ and weights $w_i = 1/N$, $i=1\ldots N$ resulting from the minimization of $J$ thus constitute GFO in its final form, and the powder average is then estimated from Eq.~(\ref{eq:Sbar_discrete}). 

Analogously, the direct extension for  the $S^l_{mn}$ required for the higher order rotational invariants would be
\begin{align*}
   S^l_{mn} = \int_{SO(3)}\mathcal{D}^{l}_{mn}(h)S(h) \mathrm{d}h,
\end{align*}
and its estimate
\begin{align*}
   \widehat{S}^l_{mn}(g) =  \sum_{i=1}^N w_i \mathcal{D}^{l}_{mn}(gh_i^{-1}) S(gh_i^{-1}).
\end{align*}
However, this form does not fully exploit the $D_2$ symmetry because of the $\mathcal{D}^l$ for which $\mathcal{D}^l(gK)\neq\mathcal{D}^l(g)$ in general. We therefore rewrite the expression so only the $D_2$ invariant part of the basis contributes, by averaging over $K\in D_2$, i.e., 
\begin{align*}
   S^l_{mn} &= \int_{SO(3)}\mathcal{D}^{l}_{mn}(h)S(h) \mathrm{d}h \\
   &= \frac14 \sum_{K\in D_2}\int_{SO(3)}\mathcal{D}^{l}_{mn}(h)S(hK) \mathrm{d}h\\
   &= \frac14 \sum_{K\in D_2}\int_{SO(3)}\mathcal{D}^{l}_{mn}(hK)S(h) \mathrm{d}h\\
   &= \int_{SO(3)}[\mathcal{D}^{l}(h)P^l]_{mn} S(h) \mathrm{d}h,
   \end{align*}
where in the third line we changed variables $h\mapsto hK$ and $K\mapsto K^{-1}$. Finally, this leads to the estimate
\begin{align*}
   \widehat{S}^l_{mn}(g) &=\int_{SO(3)}[\mathcal{D}^{l}(gh^{-1})P^l]_{mn}S(gh^{-1})f_{D_2}(h) \mathrm{d}h\\
   &=   \sum_{i=1}^N w_i [\mathcal{D}^{l}(gh_i^{-1})P^l]_{mn} S(gh_i^{-1}).
\end{align*}
As shown in the Appendix, the functions $[\mathcal{D}^{l}(gh^{-1})P^l]_{mn}$ span the $D_2$-invariant subspace at band $l$, and therefore serve as the natural basis functions on $SO(3)/D_2$, just as the Wigner $\mathcal{D}$-functions do on $SO(3)$. 

\subsection{Electrostatic repulsion and spherical designs}
The principle of electrostatic repulsion on $SO(3)$\cite{jespersenIsotropicSamplingTensorencoded2025} can also be extended to take into account $D_2$ symmetry. To do so, we replace the (geodesic) distance $d(h_i,h_j)$ between two rotations $h_i$ and $h_j$ by its minimum  over their left-cosets, $h_iD_2$ and $h_jD_2$, i.e., $\mathrm{min}\left( d(h_iK_1,h_jK_2)\right)$, where the minimum is taken over $K_1,K_2\in D_2$. Here we will simply refer to the symmetry constrained electrostatic repulsion principle as "ESR". We note that the extension of electrostatic repulsion to quotient spaces is not unique. In the axisymmetric case, conventional antipodal repulsion for $\mathbb{S}^2/\mathbb{Z}_2$ is often implemented by adding antipodal image charges and using chord distances in the embedding space, rather than by applying a potential directly to the quotient geodesic distance. Analogously, for $SO(3)/D_2$, one could construct alternative image-charge energies on the $\mathbb{S}^3$ double cover using the full lifted $Q_8 = \{\pm 1, \pm i, \pm j, \pm k\}$ orbit of each point and chord distances in $\mathbb{R}^4$. Likewise, although we use an inverse-distance potential between charges, other choices of repulsive potential are possible. The ESR scheme considered here should therefore be viewed as a simple symmetry-aware repulsion baseline, rather than as an exhaustive optimization over all possible repulsion-based designs.

Another natural scheme for selecting orientations is \emph{spherical designs}, which here are quadrature points on $SO(3)$ designed to integrate polynomials up to some degree exactly. Specifically, a spherical ``t-design'' on $SO(3)$ is a set of points $h_i\in SO(3)$, $i=1\ldots N$, that constitute an equal weight integration scheme such that
\begin{equation}
\label{eq:quad}
    \int_{SO(3)} \mathcal{D}_{mn}^l(h)\,\mathrm{d}h =\frac1{N} \sum_{i=1}^N \mathcal{D}^l_{mn}(h_i)
\end{equation}
holds exactly for all $l=0\ldots t$ and all $m,n\in -l,-l+1,\ldots l$. Since the $\mathcal{D}^l$ are a complete set, the quadrature is exact for any polynomial up to degree $t$. Spherical t-designs on $SO(3)$ were previously developed and applied for DDE\cite{jespersenOrientationallyInvariantMetrics2013}. Equation~(\ref{eq:quad}), which is complex, thus provides a total of
\[
2\sum_{l=1}^t (2l+1)^2 = \frac{2}{3} t \left(4 t^2+12 t+11\right),
\]
real constraints, while each $h_i$ provides three degrees of freedom (e.g., Euler angles). Hence a solution, which could be found numerically, should exist if $3N \geq 2/3(1+t)(1+2t)(2+3t)$. However,
as we just discussed, the relevant space is not $SO(3)$, but rather the smaller quotient space $SO(3)/D_2$. We therefore use basis functions $\Phi^l_{mn}$  of $\operatorname{Im}P^l$ in place of $\mathcal{D}^l_{mn}$ in Eq.~(\ref{eq:quad}), as described in the Appendix. Because of the smaller dimensionality $d_l$ of $\operatorname{Im}P^l$, we need fewer points to integrate up to the same degree $t$. Specifically, we find that a solution now exists when $3N\geq 1/6 \left(4 t^3+12 t^2+11 t+9 (-1)^t (t+1)-9\right)$. Choosing the smallest integer $N$ fulfilling this condition for a given $t$, we find spherical t-designs for $(t,N) = \{(2,7),(3,12),(4,30),(5,44),\ldots\}$. 

\section{Methods}
We perform minimization of Eq.~(\ref{eq:opt3}) over rotations parameterized in terms of quaternions. The maximum $l$ value used for the filter was $L = 8$. A global optimization algorithm "particleswarm" (Matlab, version \textit{R2025b}, The MathWorks, Inc., Natick, Massachusetts, United States) was used. For choosing the weightings $V_{ll}$, we use a practical generic prior based on  signals generated from random diffusion tensors for a representative triaxial 
B-tensor shape,  $\mathbb{B}=\text{diag}(0, 1/3, 2/3)$ \unit{ms/\micro m^2}.  The diffusion tensors were taken as  $\mathrm{D}=\mathrm{diag}(d_1,d_2,d_3)$ with $d_3 = 2$ \unit{\micro\meter^2/\ms} and $d_1$ and $d_2$ uniformly distributed in $[0,2]$ \unit{\micro\meter^2/\ms}.  The resulting optimization is not intended to be universally optimal across all possible signal classes, but rather to provide a robust design for a broad and relevant class of signals with axisymmetric and triaxial encoding.
We quantified the performance of the powder averages in terms of the coefficient of variation (CV) of powder-averaged signals; the ideal powder average is stationary under rotations of the object meaning that a lower CV is better. Assuming a signal from Gaussian diffusion (Eq.~(\ref{eq:Gauss}))
we calculated CV across 1183 rotations (from an $L = 6$ Euler grid, see below) of the diffusion tensor, $\mathrm{D}=\text{diag}(0.1, 0.1, 2.8)$ \unit{\micro m^2/ms}, for sets of 4 to 64 rotations of the b-tensor $\mathbb{B}=\text{diag}(0, 1/3, 2/3)$ \unit{ms/\micro m^2}.
Six schemes for generating the rotation sets were compared: (i) Haar-random rotations; (ii) quasi-uniform grid; (iii) electrostatic repulsion (ESR); (iv) GFO; (v) a "naive scheme"; and (vi) t-designs. The quasi-uniform grid was constructed on the basis of Hopf fibration coordinates in $SO(3)$ as described in \cite{yershovaGeneratingUniformIncremental2010}. 
The naive scheme was constructed by first selecting rotation axes using the principle of electrostatic repulsion on the 2-sphere \cite{jonesOptimalStrategiesMeasuring1999}, followed by picking rotation angles $\varphi$ as deterministic quantiles of the cumulative distribution $(\varphi-\sin\varphi)/\pi$ and randomly pairing them with the axes.  The normalized Haar measure in rotation-axis ($\hat{n}\in \mathbb{S}^2$) and rotation-angle ($\varphi\in[0,\pi]$) coordinates is
\[
dh=\frac{1}{4\pi^2}\sin^2(\varphi/2)\,d\varphi\,d\hat{n}.
\]
Hence, the marginal density of $\varphi$ is $p(\varphi)=\pi^{-1}\sin^2(\varphi/2)$, with cumulative distribution
\[
\int_0^\varphi p(\phi)\,d\phi=\frac{\varphi-\sin\varphi}{\pi}.
\]
Thus, the scheme reproduces the Haar angle distribution at the level of local point density, but it is not Haar-random on $SO(3)$ because both the rotation axes and the rotation angles are taken from regularized sets rather than sampled independently.

To examine if GFO also affords better sampling for purposes beyond powder averaging, we consider the performance across rotation schemes for higher order signal rotational invariants $S_l$ as defined in the Appendix.

When estimating a ground truth, we integrate using an Euler grid with exact quadrature on $SO(3)$ up to $L = 14$ (12,615 points), obtained as a direct product of a Gauss-Legendre grid ($\beta$) and 1D Fourier grids ($\alpha$ and $\gamma$). All simulations are performed in Matlab, and the optimization code and precomputed rotations sets are available open source at \url{https://github.com/Neurophysics-CFIN/GFO}.

\section{Results}\label{results}
\begin{figure*}[tbp]
  \centering
  % Subfigure (a)
  \begin{subfigure}[h]{0.48\linewidth}
    \centering
    \includegraphics[width=\linewidth]{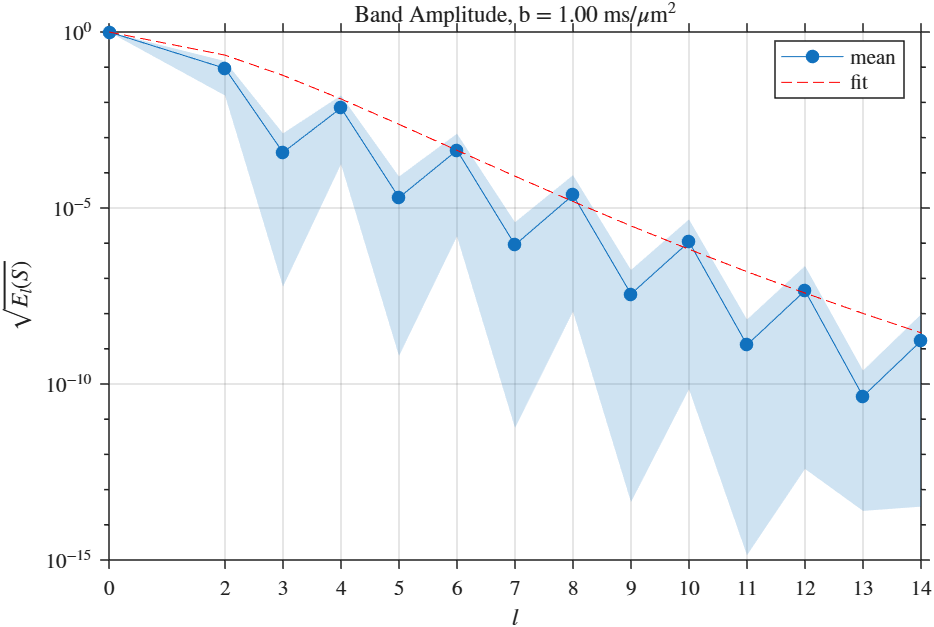}
    % \label{fig:cv_a}
  \end{subfigure}
  \hfill
  % Subfigure (b)
  \begin{subfigure}[h]{0.48\linewidth}
    \centering
    \includegraphics[width=\linewidth]{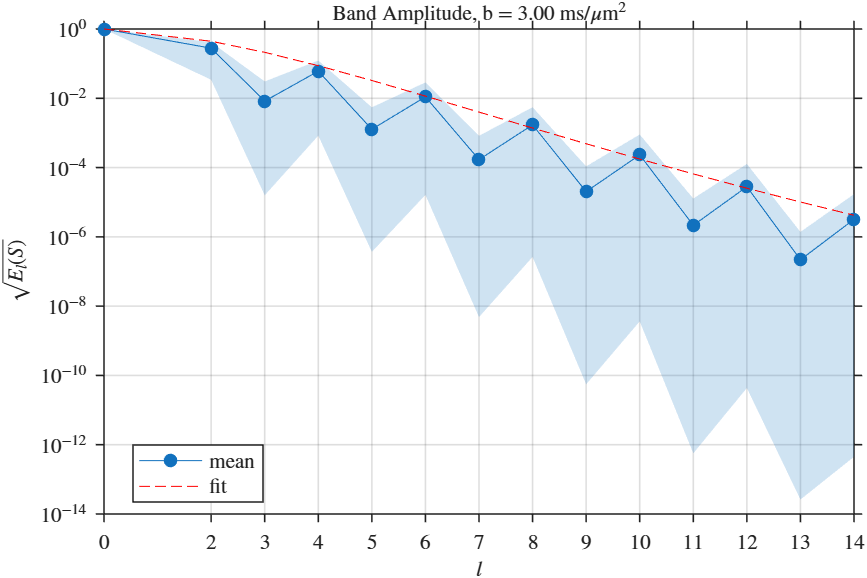}
    %\caption{Alternative scheme.}
    % \label{fig:cv_b}
  \end{subfigure}
  \caption{Relative band amplitudes as a function of $l$-band for 100 random diffusion tensors show that the power is dominated by even-$l$ bands and that power dissipates rapidly with $l$. The blue shaded region indicates the range over the distribution and circles show the mean. The red dashed line shows the Sobolev form, Eq.~\eqref{eq:sobolev}, fit to the data for even $l$, giving $\kappa \simeq 6.3$ and $s\simeq 10.7$ for $b = 1 $ \unit{\micro\meter^2/\milli\second} (a), and $\kappa \simeq 7.4$ and $s\simeq 7.9$ for $b = 3 $ \unit{\micro\meter^2/\milli\second} (b).}
  \label{fig:bandpw}
\end{figure*}

To select optimal hyper parameters ($V_{ll}$) for GFO, we plot the band amplitude $\sqrt{E_l(S)}$ for $l=0-14$ and signals generated from random diffusion tensors and $\mathbb{B} = \text{diag}(0, 1/3, 2/3)$ in Fig.~\ref{fig:bandpw}. As expected, the signal energy is dominant at even $l$, and energy at $l=1$ is exactly 0. Thus, for simplicity, our filter design neglects contributions from odd $l$. The band amplitude decays rapidly with $l$, but the mean is well described by the Sobolev form 
\begin{equation}
\label{eq:sobolev}
    \sqrt{E_l(S)}\sim (1+(l(l+1))/\kappa^2)^{-s}
\end{equation}
with $\kappa$ and $s$ values depending on $b$ such that higher b-values increase the power at higher $l$. Here, the Sobolev form is a heuristic signal prior, motivated by its stronger penalization of higher Fourier modes (larger $l$) and thus by its role as a geometrically natural smoothness prior. Since it also fits the observed decay of  $\sqrt{E_l(S)}$ well, it provides a practical proxy for the optimal variance weighting.
\begin{figure*}[tbp]
  \centering
  \begin{subfigure}[h]{0.48\linewidth}
  \includegraphics[width=\linewidth]{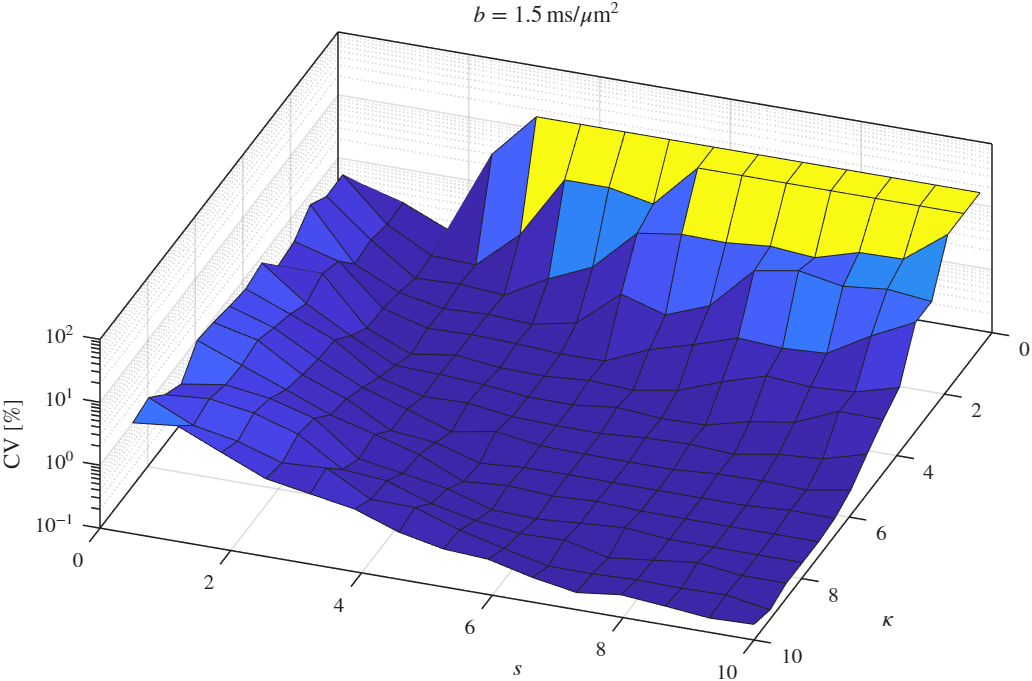}
  \end{subfigure}
  \begin{subfigure}[h]{0.48\linewidth}
  \includegraphics[width=\linewidth]{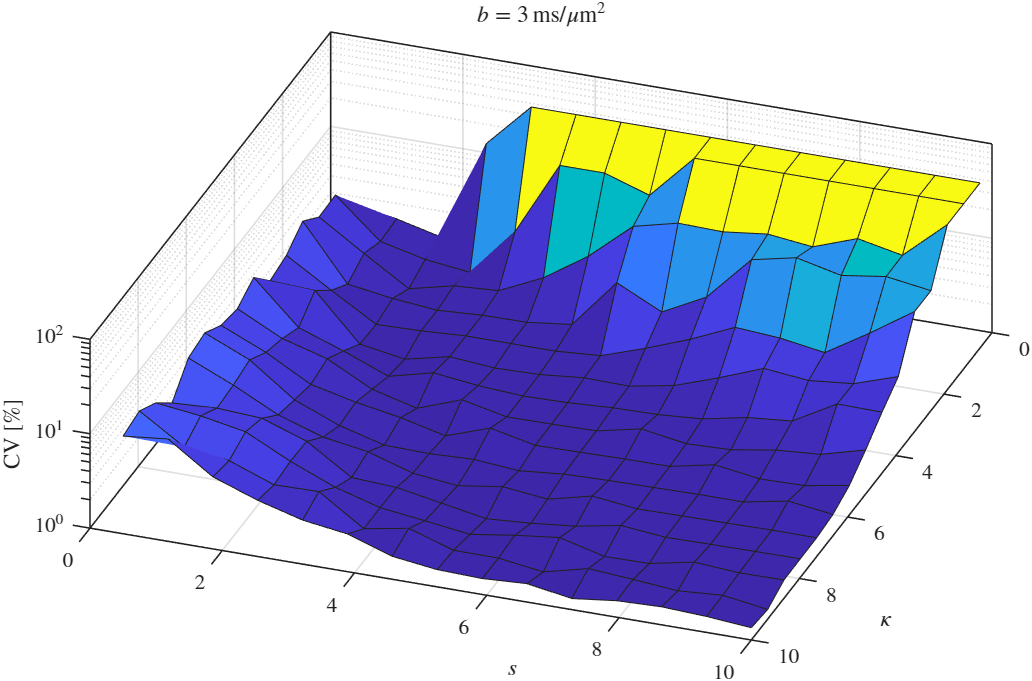}
  \end{subfigure}
\caption{The impact of hyper parameters $s$ and $\kappa$ in the GFO optimization is visualized in terms of the coefficient of variation (CV) of the signal powder average over 2601 $L=6$ Euler grid rotations of the diffusion tensor. The underlying GFO set consisted of 24 rotations and was designed with $L = 8$. Two different b-values ($b=1.5$ \unit{\micro m^2/ms} left and $b=3$ \unit{\micro m^2/ms} right) are shown.}
  \label{fig:cv_vs_hyp} 
\end{figure*}
Figure~(\ref{fig:cv_vs_hyp}) shows the effect of GFO hyper parameters $s$ and $\kappa$ on the coefficient of variation (CV) of powder average for an example diffusion and b-tensor at two b-values. Outside the low $\kappa$ or low $s$ regions, the CV is relatively flat, and this observation was consistent for other choices of B and D (not shown), suggesting that the optimization is not overly sensitive to the precise prior used to define $V_{ll}$. The main effect of increasing $b$ is to increase the overall CV. We will henceforth use  $\kappa = 7$ and $s = 8$, comfortably in the flat region of the plots.
\begin{figure*}[tbp]
  \centering
    \includegraphics[width=0.9\linewidth]{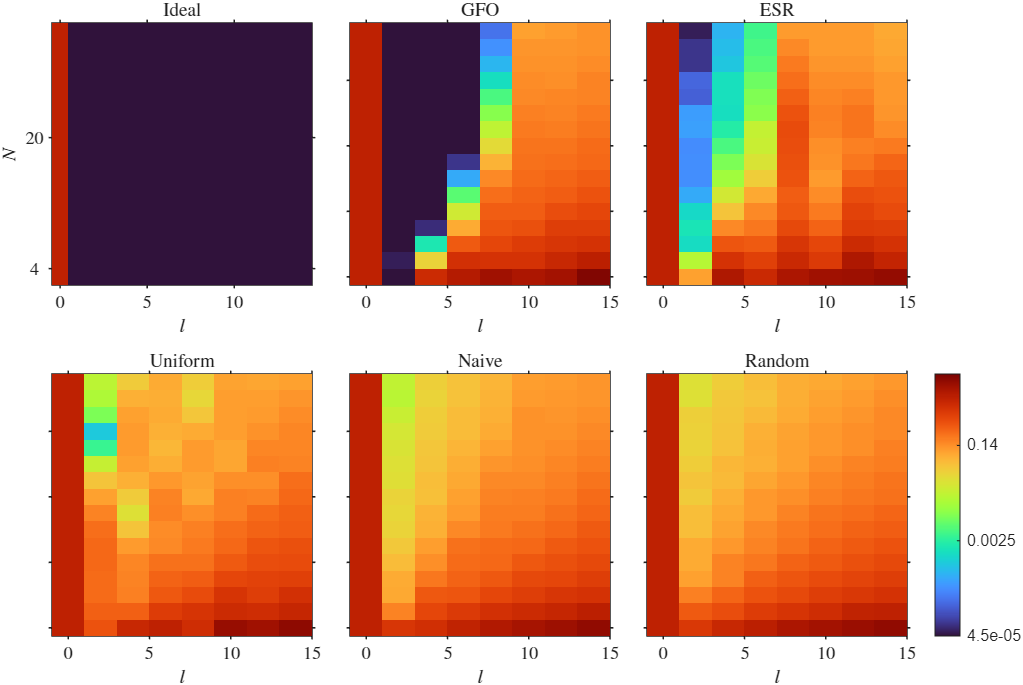}
  \caption{The spectral profiles of the GFO filters show the best approximations to the ideal. The filter band powers $E_l(f)$ across $l$ for different $N$ and all schemes. The ideal filter has $E_l(f) = \delta_{l0}$ and is shown left most. Note the logarithmic color scale.}
  \label{fig:filter_spect}
\end{figure*}

The performance of the rotation sets can be illustrated by their band powers $E_l(f)$, shown in Fig.~\ref{fig:filter_spect}. The ideal filter would have $E_l(f_0) = \delta_{l0}$ (Fig.~\ref{fig:filter_spect}, top left plot). All sets have $E_0(f) = 1$ by normalization, but nonzero $E_l(f)$ for $l>0$. This "spectral leakage" is generally highest for low $N$, and increases for increasing $l$ as expected. The GFO has the smallest leakage for $l=2$ and $l=4$, where most signal power is concentrated (cf.\ Fig.~\ref{fig:bandpw}). Both ESR and GFO improve markedly when accounting for the dihedral symmetry of the signal (data not shown).
\begin{figure*}[tbp]
  \centering
  \includegraphics[width=\linewidth]{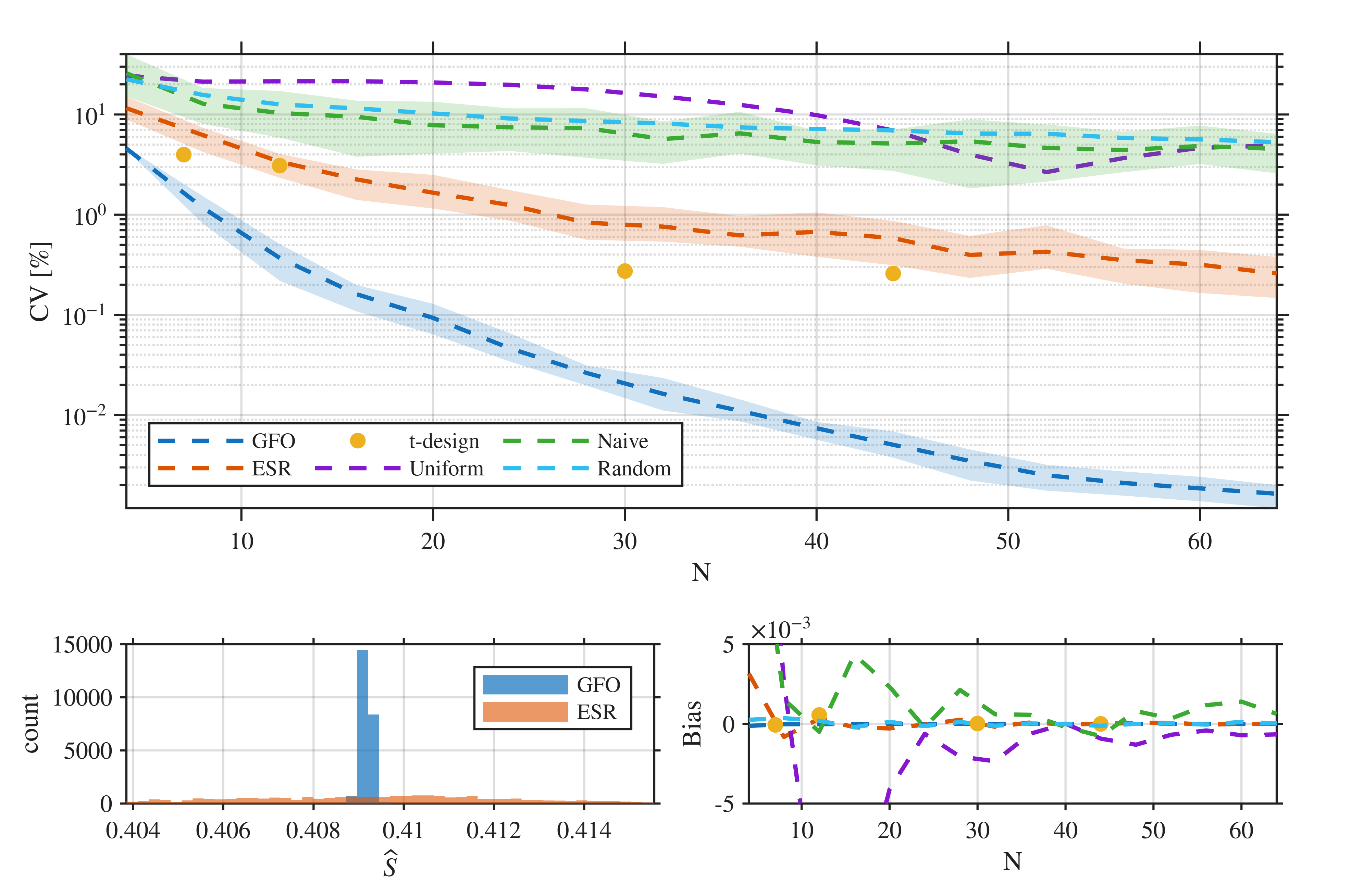}
  % \begin{subfigure}[h]{0.33\linewidth}
  % \includegraphics[width=\linewidth]{figs/CV.png}
  % \end{subfigure}
  % \begin{subfigure}[h]{0.33\linewidth}
  % \includegraphics[width=\linewidth]{figs/histo.png}
  % \end{subfigure}
  % \begin{subfigure}[h]{0.33\linewidth}
  % \includegraphics[width=\linewidth]{figs/bias.png}
  % \end{subfigure} 
  \caption{The top plot shows that GFO has the best performance across the six methods in terms of the coefficient of variation (CV) in the signal powder average (lower is better). Bottom left plot shows a representative histogram of powder averages at $N=28$ wherein GFO produces a tighter distribution compared to electrostatic repulsion. The bottom right plot shows that all methods have negligible signal bias, but electrostatic repulsion and uniform sampling have the worst overall accuracy. }
  \label{fig:cv_vs_N} 
\end{figure*}
Next, in Fig.~\ref{fig:cv_vs_N}, we present the CV for a single diffusion tensor $\mathrm{D}=\mathrm{diag}(0.1, 0.1, 2.8)$ probed with $\mathbb{B} = \mathrm{diag}(0,1/3,2/3)$ in reciprocal units, as a function of the number of directions for all considered rotation schemes. The GFO scheme exhibited the best performance with the lowest CV by a large margin. The CV from ESR and t-designs were similar and quite a bit higher than GFO, but better than the quasi-uniform, naive and random schemes which had the worst performances. The benefit of GFO is exemplified in the lower left of Fig.~\ref{fig:cv_vs_N}, where it is seen to achieve a markedly narrower distribution of signal powder averages compared to electrostatic repulsion. The bias of all schemes, shown on the lower right, is very low, but overall largest for naive and uniform sets.

\begin{figure*}[tbp]
  \centering
    \includegraphics[width=0.9\linewidth]{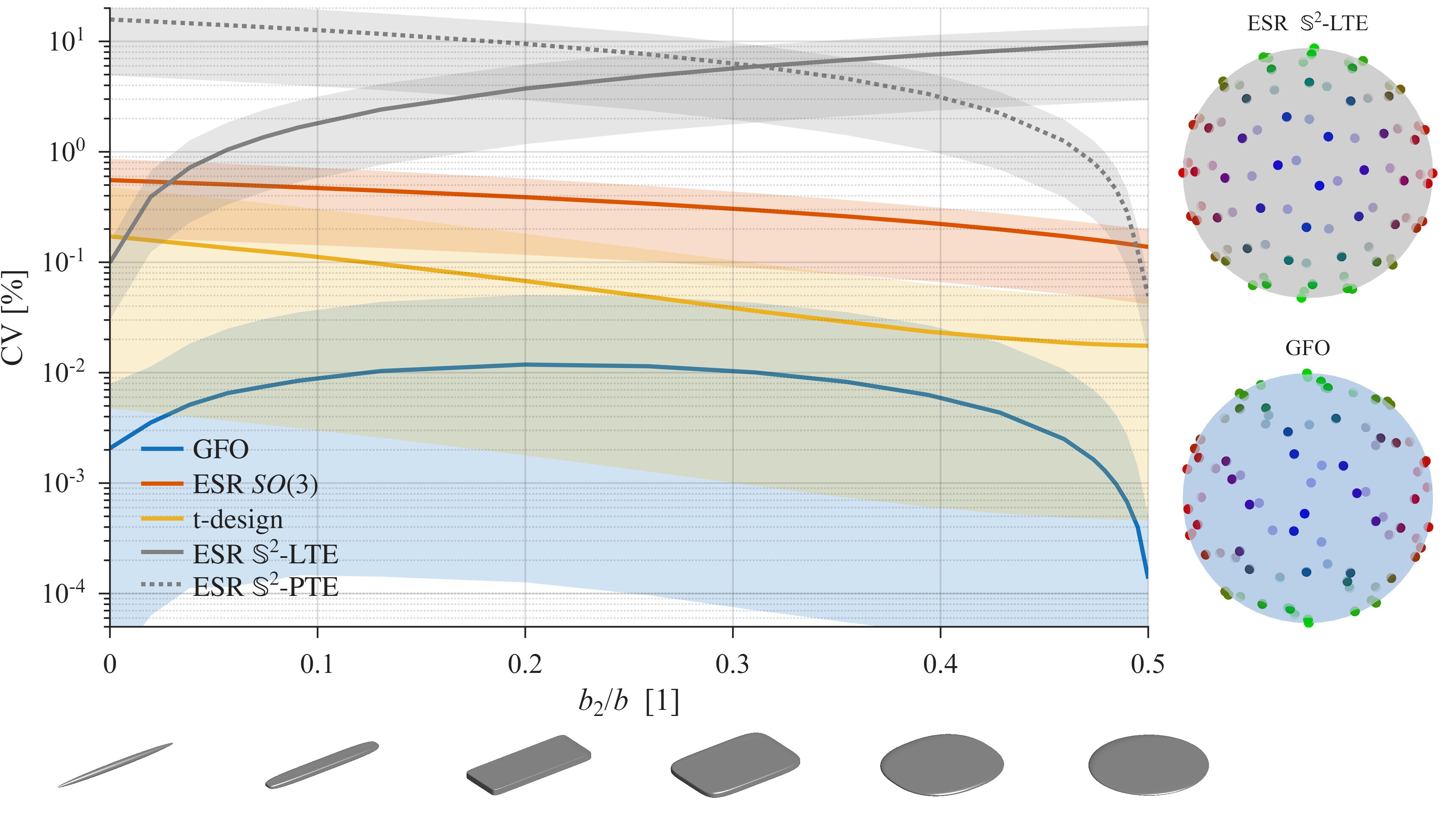}
  \caption{The CV of the signal powder average calculated from 44 rotations of the b-tensor given random rotations of a comprehensive set of diffusion tensors. Here, the shape of the b-tensor is controlled by the intermediate eigenvalue fraction $b_2/b$, as illustrated by the corresponding glyphs. The shaded bands cover the 5-95\% percentiles of CVs, and the solid line shows the median. In this case, GFO is always best, even for axisymmetric linear ($b_2/b=0$) and planar b-tensor encoding ($b_2/b=0.5$). The plots on the right show the directions of b-tensor symmetry axes that result from GFO and conventional ESR when applied to LTE. Notably, GFO outperforms conventional ESR despite its non-uniform distribution on the sphere.}
  \label{fig:cv_allD}
\end{figure*}

To find if GFO is widely applicable across b-tensor shapes, its performance was evaluated for a wide range of b-tensor shapes, spanning the edge connecting linear to planar b-tensors in Fig.~\ref{fig:B}. For reference, we compared GFO, t-design, ESR on $SO(3)$, as well as ESR on $\mathbb{S}^2$. The latter is the current default for linear and planar b-tensor encoding, wherein the symmetry axis of the b-tensor is directed along a set of points on the sphere\cite{jonesOptimalStrategiesMeasuring1999}. Fig.~\ref{fig:cv_allD} shows the CV of the signal powder average calculated across 1600 rotations of diffusion tensors (as in Fig.~\ref{fig:bandpw}) as a function of b-tensor shapes using \(N=44\) rotations. We observe that GFO performs best for the \textit{entire range} of shapes considered. Even if the distributions of CV overlap, GFO is always best when comparing pair-wise performance for a given combination of B and D. Remarkably, GFO also outperforms all other methods for axisymmetric b-tensors, including conventional ESR on $\mathbb{S}^2$, despite yielding a distribution of points on the sphere that is appreciably less uniform.

\begin{figure*}[tbp]
  \centering
  \includegraphics[width=\linewidth]{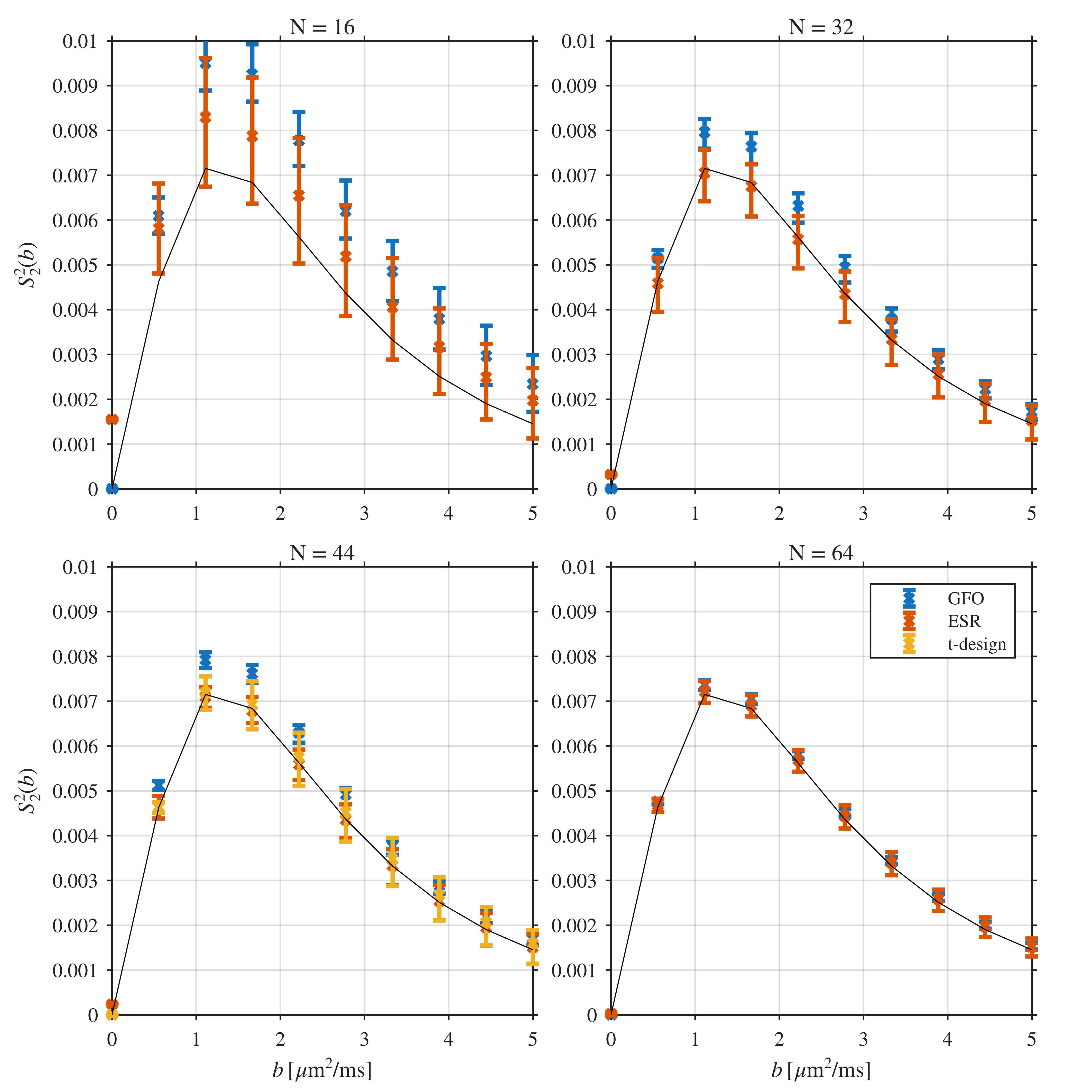}
  \caption{The precision and accuracy of higher order rotational invariants also benefit from optimized sets. The figure compares $S_2^2$ for  16, 32, 44  and 64 directions for GFO and electrostatic repulsion variants. Ground truth is shown as solid black line, underlying signal generated with $\mathbb{B} = \mathrm{diag}(0,1/3,2/3)$ and $\mathbb{D} = \mathrm{diag}(0.1,0.1,2.8)$ in reciprocal units. The t-design is only shown for the value of $N=44$, as the corresponding quotient-space design was not constructed for the other $N$.}
  \label{fig:S_2} 
\end{figure*}

We also investigated whether GFO benefits other quantities that require rotated sampling. Specifically, we evaluated the first non-vanishing higher-order rotational invariant $S_2^2(b)$ as a function of $b$ for all schemes with various $N$ (Fig.~\ref{fig:S_2}). Throughout, GFO had superior precision (lower variability) compared to both electrostatic repulsion and the t-design. Notably, biases were always positive owing to being squares of unbiased estimates $S^l_{mn}(b)$. A similar effect holds for LTE higher order invariants $S_l$, $l>0$, partially explaining the poorer fits based on such features\cite{parisThermalNoiseLowers2026}, as in RotInv\cite{novikovRotationallyinvariantMappingScalar2018}. As expected, accuracy generally improves with larger $N$, but except for $N =64$ where GFO and ESR were very similar, GFO had the largest bias. This illustrates that optimizing for isotropy in $l = 0$ features does not automatically optimize for $l = 2$ features and above.

\begin{figure*}[tbp]
  \centering
  \includegraphics[width=\linewidth]{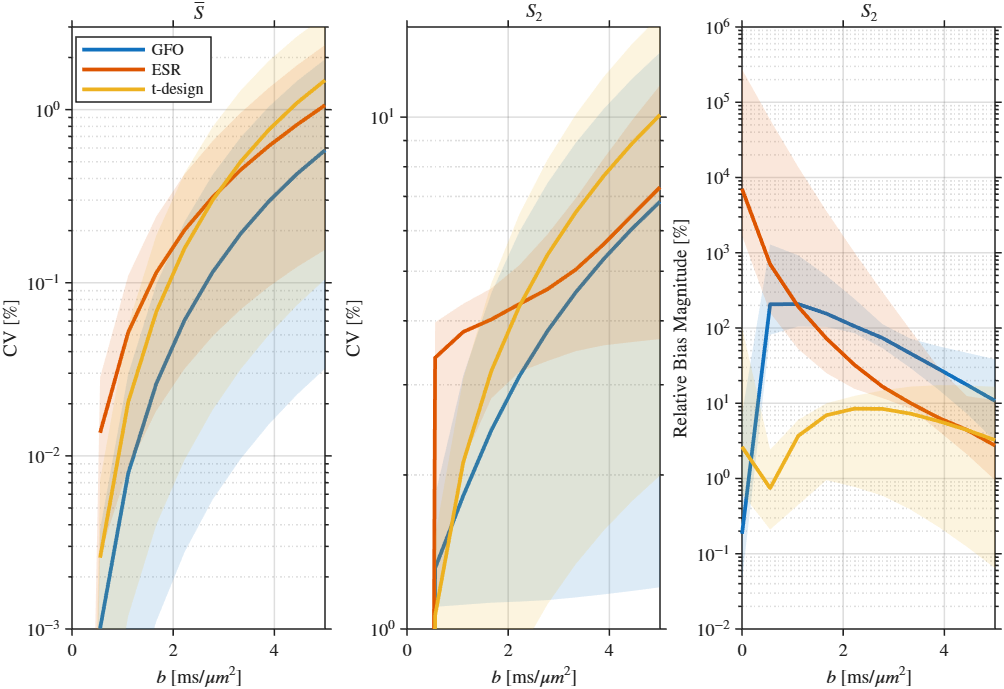}
  \caption{The precision and accuracy of $\bar{S}$ and $S_2$ for 3 orientation sets considered over the distribution of diffusion tensors. The CV for each diffusion tensor was averaged over the distribution and plotted for $\bar{S}$ and $S_2$ in left and middle plots, respectively. The bias was defined as the absolute value of the estimate minus the ground truth and normalized to the peak ground truth amplitude, and subsequently averaged over the diffusion tensor distribution. It is shown for $S_2$ in right plot, but not shown for $\bar{S}$, for which it is very small, c.f.\ Fig.~\ref{fig:cv_vs_N}. In all graphs, lines are the medians and shaded region covers the 5--95\% range. The diffusion b-tensor shape was  $\mathbb{B} = \mathrm{diag}(0,1/3,2/3)$  \unit{ms/\micro\meter^2}.}
  \label{fig:S0S2}
\end{figure*}

Finally, we investigated the behavior of GFO, ESR and the t-design estimates of both $\bar{S}$ and $S_2$ for $N = 44$ in Fig.~\ref{fig:S0S2}. The previously observed superiority of GFO in the coefficient of variation of the powder average persists across both b-values and diffusion tensors. For the $S_2(b)$, the best scheme in terms of CV depends on $b$; the t-design generally has the lowest CV for the lowest $b$-values, whereas GFO performed best for $b\gtrsim 1$  \unit{ms/\micro\meter^2}.  Regarding the bias of $S_2$, the t-design generally has the smallest bias except for  very low  where GFO wins, or very high $b$, where ESR wins. GFO  has the highest bias in $S_2$ over most of the $b$-range examined. Note that the relative biases are high, but the $S_2$ values are also quite small, c.f.\ Fig.~\ref{fig:S_2}.
\section{Discussion and conclusions}\label{discussion}
This work addresses a largely overlooked problem in diffusion MRI: how to design rotation sets that yield accurate orientational sampling when the diffusion encoding does not have a symmetry axis. While it is well established that directions for linear b-tensor encoding can be optimized to be equidistant on the surface of a sphere \cite{bakREPULSIONNovelApproach1997,jonesOptimalStrategiesMeasuring1999}, the general case of triaxial b-tensor shapes has previously been viewed as requiring uniform sampling over the rotation group $SO(3)$, for which fewer practical solutions exist \cite{jespersenOrientationallyInvariantMetrics2013,jespersenIsotropicSamplingTensorencoded2025,Jespersen13,westinPlatonics2020}.  Indeed, we have shown that a naive application of existing rotation schemes to triaxial encoding is ill-advised as it can lead to substantial bias and variability in powder-averaged signals. We solve this problem by GFO, which performs well with axisymmetric encoding, and is categorically superior to the other methods considered here for triaxial encoding. Therefore, we expect GFO to become increasingly valuable as non-conventional encoding strategies are adopted.

A central conceptual result of this work is the identification of an intrinsic dihedral symmetry of diffusion signals generated by arbitrary tensor-valued encoding. As a consequence, the relevant signal space is not the full rotation group $SO(3)$ but the quotient space $SO(3)/D_2$. This symmetry is important because it establishes the correct geometric setting for harmonic analysis and rotational sampling, and thus also the way rotation sets should be designed for non-axisymmetric encoding. The contribution of the present work is therefore twofold: it clarifies the symmetry structure of the problem, and it uses that structure to formulate sampling schemes directly on the appropriate quotient space. Within this common framework, the electrostatic repulsion, spherical t-design, and GFO schemes arise as natural symmetry-aware constructions.

Our specific methodological contribution within this common quotient-space framework is GFO: a filter-based strategy for constructing rotation sets that approximate the ideal powder-averaging filter while emphasizing the harmonics most relevant for powder averaging. Rather than seeking a generic notion of uniformity, GFO uses prior information about the signal band powers to reduce unwanted filter power in the low-order components that dominate the rotation dependence of the powder-average estimate. In that sense, the method is adapted not only to the geometry of the sampling space, but also to the structure of the signal class of interest. The $D_2$ reformulation further sharpens this construction by restricting attention to the symmetry-allowed part of the spectrum, thereby improving the efficiency of the optimization. At the same time, this symmetry-aware viewpoint does not benefit GFO alone, but also provides a fairer and more natural basis for comparing alternative schemes such as ESR and spherical t-designs on the same quotient space.

The main practical finding was that, for powder averaging, GFO provides the best overall performance among the schemes considered. Across the tested rotation sets, it consistently yielded the lowest coefficient of variation together with negligible bias, showing that the filter-based design can be translated into a tangible gain in sampling efficiency. Importantly, this advantage was not confined to triaxial b-tensors, but rather general across b-tensor shapes. Across the range of b-tensor shapes examined, GFO performed best for essentially the entire interval considered, indicating that its benefit extends broadly across non-axisymmetric encoding and is not restricted to a narrow region of parameter space. Notably, even in the axially symmetric case, it outperformed standard electrostatic repulsion. In practical terms, this means that more accurate powder averages can be obtained with fewer orientations, or conversely, that a given level of performance can be achieved at shorter scan time.

The results for the higher-order rotational invariant $S_2$ were more nuanced. The same quotient-space framework naturally supports the estimation of such invariants, but the GFO optimization was specifically designed for powder averaging, i.e., the $l=0$ sector, and therefore does not in itself guarantee optimal performance for $l\geq 2$. Consistent with this, no single scheme emerged as uniformly best across both bias and precision for $S_2$ over the range of $b$-values examined. Rather, GFO, ESR, and spherical t-designs exhibited different trade-offs depending on $b$ and $N$. This highlights that accurate estimation of higher-order invariants is a related but distinct sampling problem, and suggests that schemes optimized specifically for these quantities may be needed when they are the primary target of the experiment. This remains relevant for rotational-invariant analyses and model-based approaches that rely on such quantities. A relevant recent example is the RICE framework\cite{coelhoGeometryCumulantSeries2026}, which explicitly computes $SO(3)$ Fourier components and corresponding rotational invariants of the diffusion signal under arbitrary tensor encoding. Similarly, recent work on non-axisymmetric fiber bundles in white matter motivates extending the Standard Model to a convolution on $SO(3)$, further emphasizing a role for rotation-set designs beyond scalar powder averaging \cite{CoelhoISMRM2026}. 

Our findings have broader implications for diffusion MRI methods that rely on accurate orientational sampling under general tensor-valued encoding. In particular, the framework is well suited for techniques that employ triaxial encoding tensors, such as skewness tensor imaging \cite{ningProbingTissueMicrostructure2021}. More generally, the same geometric viewpoint should also be relevant for encoding features beyond the b-tensor itself, such as orientation-dependent diffusion-time effects in restriction-weighted q-space trajectory imaging \cite{szczepankiewiczRestrictionweightedQspaceTrajectory2025}. Because powder averages also play an important role in model-based analyses, improved rotational sampling may likewise benefit advanced microstructural models and protocols that combine multiple b-tensor shapes, including the Standard Model \cite{novikovQuantifyingBrainMicrostructure2019} and its exchange extensions, SMEX/NEXI \cite{olesenDiffusionTimeDependence2022,jelescuNeuriteExchangeImaging2022}, as well as correlation tensor imaging \cite{henriquesCorrelationTensorMagnetic2020,henriquesDoubleDiffusionEncoding2021}.

Several limitations should be noted. First, the present GFO optimization was targeted specifically to powder averaging, and thus to the $l=0$ sector, so optimal performance for higher-order invariants is not guaranteed. As discussed above for $S_2$, this may lead to trade-offs relative to schemes that perform better for particular combinations of $b$ and $N$. Second, all simulations were performed for noise-free signals assuming Gaussian diffusion. This isolates orientation-induced effects, but future work should assess robustness under realistic noise and more elaborate signal models. Third, although the performance of GFO was favorable across the range of b-tensor shapes examined, the exploration of shape space was still limited, and a fuller mapping of performance across acquisition settings remains warranted. 
Fourth, GFO relies on a prior model for the signal angular power spectrum. Improved rotation sets may therefore be achievable when more specific prior information about the signal class is available. In a preliminary analysis \cite{JespersenISMRM2026}, we also examined whether post-hoc measurement weighting, analogous to Knutsson weighting for LTE \cite{knutssonAdvancedFilterDesign1999,afzaliComputingOrientationalaverageDiffusionweighted2021,szczepankiewiczMeasurementWeightingScheme2017}, could further improve powder averaging for GFO and other reasonably uniform rotation sets. However, we observed no meaningful improvement, suggesting that for well-designed sets, the residual low-order structure is already sufficiently suppressed that uniform weights are adequate. An interesting extension of the present framework would instead be to reverse the procedure and compute optimal weights for existing rotation sets, which could retrospectively improve powder averages in previously acquired data. Finally, very high $b$-values may require extending the harmonic bandwidth $L$ used in the optimization. These considerations do not, however, pose a practical barrier to the use of GFO, since the method requires only a one-time offline optimization and the computational cost was negligible for the rotation-set sizes considered here, including $N=64$.

In conclusion, we identified an intrinsic $D_2$ symmetry of diffusion signals under arbitrary tensor-valued encoding, showing that their natural rotational signal space is the quotient space $SO(3)/D_2$. This provides the appropriate geometric framework for harmonic analysis and symmetry-aware design of rotation sets for non-axisymmetric encoding. Within this framework, GFO provided the strongest overall performance for powder averaging across the b-tensor shapes examined, while higher-order invariants such as $S_2$ were shown to constitute a related but distinct optimization target. More broadly, these results support the use of symmetry-adapted rotational sampling as a general principle for multidimensional diffusion MRI.

\section*{Acknowledgments}
The authors are grateful to Dmitry Novikov for discussions and substantial input that significantly strengthened the paper. The authors also thank Noam Shemesh and Santiago Coelho for valuable discussions.
 This work was partially supported by Lundbeck Foundation grant 10.46540/3103-00144B, the Swedish Cancer Society grant 22 0592 JIA, and the Crafoord Foundation grant 20240791. SNJ  is also grateful for support from The Danish Research Foundation. Generative AI tools, specifically ChatGPT, were utilized in this research to assist in code and text drafting. The AI outputs have been rigorously verified for accuracy, and their use has been disclosed in compliance with the University's guidelines. The researchers take full responsibility for all outputs.

\subsection*{Financial disclosure}
None reported.

\subsection*{Conflict of interest}
The authors declare no potential conflict of interests.

%\clearpage
\section{Appendix}
\subsection{Cost function}

Here we derive the cost function in Eq.~(\ref{eq:opt}) by first writing out the product 
\begin{multline}
\label{eq:J_app}
J(\{h_i\}_{i=1\ldots N})=(\mathrm{A}\mathbf{w}-\mathbf{f}_0)^\dagger \mathrm{V}^\dagger \mathrm{V} (\mathrm{A}\mathbf{w}-\mathbf{f}_0)=\\
(\mathrm{A}\mathbf{w})^\dagger \mathrm{V}^\dagger \mathrm{V} (\mathrm{A}\mathbf{w}) + \mathbf{f}_0^\dagger\mathrm{V}^\dagger \mathrm{V} \mathbf{f}_0 - \mathbf{f}_0^\dagger\mathrm{V}^\dagger \mathrm{V} (\mathrm{A}\mathbf{w}) - \\
(\mathrm{A}\mathbf{w} )^\dagger \mathrm{V}^\dagger \mathrm{V}\mathbf{f}_0.
\end{multline}
We next consider the terms individually, but first refresh the definitions of the variables involved. The matrix A has size $N_L\times N$ and contains the Wigner-D matrices $\mathcal{D}^l_{mn}(h_i)$ with the composite index $lmn$ running over its rows (up to some maximal $l=L$), and $i$ running over its columns. The $N_L\times 1$ vector  $\mathbf{f}_0$ stacks the Fourier coefficients of the target filter,  $(f_0)^l_{mn} = \delta_{l0}\delta_{m0}\delta_{n0}$, and $\mathbf{w}$, also $N_L\times 1$,  are the weights of each orientation in the powder average. Finally, the matrix V is diagonal and is a lever for modulating the contribution of different $l$-bands. We take all variables except A to be strictly real.

The term $\mathbf{f}_0^\dagger\mathrm{V}^\dagger \mathrm{V} \mathbf{f}_0$ in Eq.~(\ref{eq:J_app}) does not involve the optimization variables $h_i$ and can be regarded as a constant. The same goes for the mixed terms, $\mathbf{f}_0^\dagger\mathrm{V}^\dagger \mathrm{V} (\mathrm{A}\mathbf{w}) + (\mathrm{A}\mathbf{w} )^\dagger \mathrm{V}^\dagger \mathrm{V}\mathbf{f}_0$, because V is diagonal and $\mathbf{f}_0$ non-zero only in the first component ($l=m=n=0$), where A is constant, e.g.:
\begin{multline}
    (\mathrm{A}\mathbf{w} )^\dagger \mathrm{V}^\dagger \mathrm{V}\mathbf{f}_0 =  \sum_{ij}A_{ij}^*w_j V_{ii}^2 f_{0i} = \\
    \sum_{lmnj}\mathcal{D}^{l*}_{mn}(h_j) w_j V_{ll}^2 \delta_{l0}\delta_{m0}\delta_{n0} 
    = \sum_j\mathcal{D}^{0*}_{00}(h_j) w_j V_{00}^2 =  V_{00}^2,
\end{multline}
since $\mathcal{D}^0_{00}\equiv 1$ and $\sum_jw_j=1$. For the last term, we find using $w_j =1/N$ 
\begin{multline}
(\mathrm{A}\mathbf{w})^\dagger \mathrm{V}^\dagger \mathrm{V}(\mathrm{A}\mathbf{w}) =  \sum_{ijk}A_{ij}^*w_j V_{ii}^2 A_{ik}w_k\\
= \frac1{N^2}\sum_{j,k=1}^N\sum_{l=0}^L V_{ll}^2 \sum_{m,n=-l}^l \mathcal{D}^l_{mn}(h_j) \mathcal{D}^{l*}_{mn}(h_k) \\
=\frac1{N^2}\sum_{j,k=1}^N\sum_{l=0}^L V_{ll}^2 \sum_{m,n=-l}^l \mathcal{D}^l_{mn}(h_j) \mathcal{D}^l_{nm}(h_k^{-1})\\
=\frac1{N^2}\sum_{j,k=1}^N\sum_{l=0}^L V_{ll}^2 \sum_{m=-l}^l \mathcal{D}^l_{mm}(h_jh_k^{-1}) \\
=\frac1{N^2}\sum_{j,k=1}^N\sum_{l=0}^L V_{ll}^2 \Tr ( \mathcal{D}^l(h_jh_k^{-1})) \\
=  \frac1{N^2}\sum_l V_{ll}^2  \sum_{jk} \chi^l\big(h_jh_k^{-1}\big).
\end{multline}
Here we used fundamental properties of a unitary group representation, $\mathcal{D}^l(h_1)\mathcal{D}^l(h_2) = \mathcal{D}^l(h_1h_2)$ and $\mathcal{D}^l(h^{-1}) = \mathcal{D}^{l}(h)^\dagger$. Using a standard formula for the $SO(3)$ group characters $\chi^l(h) \equiv \Tr \mathcal{D}^l(h)$, and collecting the 3 constant terms in $c$, leads immediately to the expression in Eq.~(\ref{eq:opt}).

As mentioned in the main text, the resulting estimate of the powder average, $\widehat{S}(g)$, may still depend on the relative orientation $g$ of B, and we can therefore analyze its variance over $g$. For that we utilize $\widehat{S}^l = S^l f^l$, such that
\begin{align*}
\widehat{S}(g) &=\sum_{l,m,n} (2l+1)\mathcal{D}^{l*}_{mn}(g)\widehat{S}^l_{mn}\\
&= \sum_{l,m,n,n'} (2l+1)\mathcal{D}^{l*}_{mn}(g)S^l_{mn'}f^l_{n'n}.    
\end{align*}
Averaging over $g$ and using orthogonality of Wigner functions, we have immediately
\[
\langle \widehat{S}(g)\rangle = \int_{SO(3)}\widehat{S}(g)\mathrm{d}g = S^0_{00}f^0_{00} = \bar{S},
\]
such that the estimator is unbiased. For the second moment, using that the signal is real and orthogonality of Wigner functions, we get
\begin{align*}
\langle |\widehat S(g)|^2\rangle
&=
\Bigl\langle
\sum_{l_1,m_1,n_1}(2l_1+1)\mathcal D^{l_1*}_{m_1n_1}(g)\widehat S^{l_1}_{m_1n_1}\times\\
&\sum_{l_2,m_2,n_2}(2l_2+1)\mathcal D^{l_2}_{m_2n_2}(g)\widehat S^{l_2*}_{m_2n_2}
\Bigr\rangle \\
&=
\sum_{l,m,n}(2l+1)\,|\widehat S^l_{mn}|^2 \\
&=
\sum_{l,m,n}(2l+1)\left|\sum_{n'} S^l_{mn'}f^l_{n'n}\right|^2 .
\end{align*}
Computing the variance by subtracting the square of the mean removes the $l=0$ component, such that
\begin{align*}
    \operatorname{Var}(\widehat{S}(g))&=\langle (\widehat{S}(g))^2\rangle-\langle \widehat{S}(g)\rangle^2 \\
    &=  \sum_{l>0}(2l+1)\sum_{m,n}
\Bigl| \sum_{n'}
S^{l}_{mn'}f^{l}_{n'n}
\Bigr|^2 .
\end{align*}
as stated in Eq.~(\ref{eq:SO3-var-exact}) of the  main text.

\subsection{Right-\texorpdfstring{$D_2$}{D2}-invariant functions and the quotient space \texorpdfstring{$SO(3)/D_2$}{SO(3)D2}}

Here, we collect some basic facts about functions $S(g)$ on $SO(3)$ that are right-invariant under the dihedral subgroup
\[
D_2=\{I,R_x(\pi),R_y(\pi),R_z(\pi)\}\subset SO(3).
\]
In the present setting, this is the relevant symmetry for diffusion signals of the form
\[
S(g)=S(b,g\mathbb{B}g^{-1}),
\]
where $\mathbb{B}$ is the diagonal b-tensor in its principal frame. Indeed, for any $K\in D_2$,
\[
S(gK)=S(b,gK\mathbb{B}K^{-1}g^{-1})=S(b,g\mathbb{B}g^{-1})=S(g),
\]
since $K\mathbb{B}K^{-1}=\mathbb{B}$.
Because $S(g)$ is constant on the sets
\[
gD_2=\{gK:\ K\in D_2\},
\]
it is naturally a function on the quotient space
\[
SO(3)/D_2.
\]
Equivalently, functions on $SO(3)/D_2$ may be identified with functions on $SO(3)$ satisfying
\[
S(gK)=S(g),\qquad \forall g\in SO(3),\ \forall K\in D_2.
\]
Colloquially, this says that as far as the signal is concerned, the four b-tensors obtained by rotations of $\mathbb{B}$ by $gK$ for $K\in D_2$, are all equivalent.

It is useful to encode this symmetry by means of a projector defined as a function on $SO(3)$,
\[
P(g)\equiv \frac14\sum_{K\in D_2}\delta(gK^{-1}),
\]
so that, with the convolution convention of Eq.~\eqref{eq:conv}, we get that
\[
(S\otimes P)(g)
=
\int_{SO(3)} S(gh^{-1})P(h)\,dh
=
\frac14\sum_{K\in D_2} S(gK^{-1}).
\]
Since $D_2$ is closed under inversion, this becomes
\[
(S\otimes P)(g)=\frac14\sum_{K\in D_2} S(gK).
\]
Thus, convolution with $P$ is precisely averaging over the right action of $D_2$. In particular, if $S$ is right invariant to $D_2$, then $S=S\otimes P$. But the converse also holds, i.e., if $S=S\otimes P$, then $S$ is right invariant to any $K\in D_2$:
\[\begin{split}
S(gK) = (S\otimes P)(gK) = \frac14 \sum_{K'\in D_2} S(gKK') \\
= \frac14 \sum_{K''\in D_2} S(gK'') = (S\otimes P)(g) = S(g)
\end{split}\]
Hence,
\[
S=S\otimes P \quad\Leftrightarrow\quad S\text{ is right-invariant }\quad\Leftrightarrow \quad S^l = S^l P^l
\]
where we invoked the convolution theorem
\[
(S\otimes P)^l = S^l P^l
\]
for the last equivalence. Thus, working with the projector $P^l$ is fully equivalent to imposing invariance under the four individual group elements.

In Fourier space, the projector is represented by the Wigner-D matrices
\[
P^l =\frac14\sum_{K\in D_2} \mathcal{D}^l(K)=(P^l)^\dagger,
\]
which follows directly from its definition. That $P^l$ is indeed a projector is verified by
\[
(P^l)^2
=
\frac1{16}\sum_{K_1,K_2\in D_2}\mathcal{D}^l(K_1K_2)
=
\frac14\sum_{K\in D_2}\mathcal{D}^l(K)
=
P^l.
\]

The explicit form of $P^l$ follows from standard identities for the Wigner matrices, giving
\begin{align*}
\mathcal{D}^l_{mn}(I)&=\delta_{mn}\\
\mathcal{D}^l_{mn}(R_z(\pi))&=\delta_{mn}(-1)^n\\
\mathcal{D}^l_{mn}(R_y(\pi))&=(-1)^{l+n}\delta_{m,-n}\\
\mathcal{D}^l_{mn}(R_x(\pi))&=(-1)^l\delta_{m,-n}
\end{align*}
Hence,
\begin{align}
    \label{eq:projector}
&P^l_{mn}\nonumber\\
&=\frac14 \big( \delta_{mn}+ (-1)^n\delta_{mn} + (-1)^{l+n}\delta_{m,-n} + (-1)^l\delta_{m,-n}\big)\nonumber\\
&=\frac14 \big(1+(-1)^n\big)\big(
\delta_{mn}
+
(-1)^{l}\delta_{m,-n}
\big).
\end{align}

Since $S^l = S^l P^l$, each row of $S^l$ lies in the row space of $P^l$, which here coincides with the column space, i.e., $\mathrm{Im}\,P^l$. Thus, for each fixed $m$, the transposed row vector
\[
\left(S^l_{m,-l}, S^l_{m,-l+1}, \ldots, S^l_{m,l-1}, S^l_{m,l}\right)^T
\]
belongs to $\mathrm{Im}\,P^l$. Consequently, the right index $n$ does not range freely over the full $(2l+1)$-dimensional space, but only over the $D_2$-invariant subspace. In particular, from the explicit form of $P^l$ in Eq.~(\ref{eq:projector}), we can immediately infer the selection rules: because of the factor $1+(-1)^n$, only even $n$ can contribute. Moreover, the second factor in Eq.~(\ref{eq:projector}) couples the $n$ and $-n$ components, leaving the $n=0$ component for even $l$, together with normalized combinations of the $n$ and $-n$ components for even $n>0$. Consequently, 
\[
S^l_{mn}=0
\qquad\text{for odd } n,
\]
and, 
\[
S^1=0.
\]
By contrast, odd-$l$ coefficients do not vanish in general for $l>1$, although they all have $S^l_{00} = 0$, c.f. Eq.~(\ref{eq:projector}).

The dimension of $\mathrm{Im}\, P^l$ is
\[
d_l \equiv \dim(\operatorname{Im}P^l)=\operatorname{Tr}(P^l),
\]
since $P^l$ is a projector. Using the explicit form above we find 
\[
d_l=\frac14\Big[(2l+1)+3(-1)^l\Big]
=
\begin{cases}
\dfrac{l}{2}+1, & l\ \text{even},\\[2mm]
\dfrac{l-1}{2}, & l\ \text{odd}.
\end{cases}
\]
In particular,
\[
d_0=1,\quad d_1=0,\quad d_2=2,\quad d_3=1,\quad d_4=3, \ldots
\]
Thus we see again that the entire $l=1$ sector vanishes for any right-$D_2$-invariant function.

A convenient basis of $\operatorname{Im}P^l$ is obtained by choosing a matrix
\[
U^l\in\mathbb{C}^{(2l+1)\times d_l}
\]
whose columns form an orthonormal basis for $\operatorname{Im}P^l$. Then
\[
P^l=U^l(U^l)^\dagger,
\qquad
(U^l)^\dagger U^l = I_{d_l}.
\]
Expanding the rows of $S^l$ in that basis yields
\begin{equation}
\widetilde{S}^l \equiv S^lU^l,
\qquad
S^l=\widetilde{S}^l(U^l)^\dagger.
\label{eq:Stilde}
\end{equation}

Finally, this leads naturally to a basis for functions on the quotient space $SO(3)/D_2$.\footnote{An explicit expression for this basis was first found by Dmitry Novikov. In our simulations, we used SVD.} Using $S^l = S^l P^l = S^l U^l (U^l)^\dagger$ in the $SO(3)$ expansion, Eq.~(\ref{eq:FT1}), 
\[
\begin{split}
S(g)= \sum_l (2l+1)\Tr \left(S^l\mathcal{D}^l(g)^\dagger\right)\\ = \sum_l (2l+1)\Tr \left(S^lU^l (U^l)^\dagger\mathcal{D}^l(g)^\dagger\right)\\
=\sum_l (2l+1)\Tr \left(\widetilde{S}^l (\mathcal{D}^l(g)U^l)^\dagger\right)\\
=\sum_l (2l+1)\Tr \left(\widetilde{S}^l \widetilde{\mathcal{D}}^l(g)^\dagger\right),
\end{split}
\]
where
\[
\widetilde{\mathcal{D}}^l(g)\equiv \mathcal{D}^l(g)U^l\in\mathbb{C}^{(2l+1)\times d_l}.
\]
Hence any function on $SO(3)/D_2$ can be expanded in terms of the $\widetilde{\mathcal{D}}^l$, which are well defined on the quotient $SO(3)/D_2$ since for any $K\in D_2$,
\[
 \widetilde{\mathcal{D}}^l(gK) = \mathcal{D}^l(g)\mathcal{D}^l(K)U^l =\mathcal{D}^l(g)U^l=\widetilde{\mathcal{D}}^l(g),
\]
where we used that $\mathcal{D}^l(K)U^l =U^l$. The latter follows since the columns of $U^l$ lie in $\mathrm{Im}\,P^l$, i.e., 
\[U^l = P^l U^l \Rightarrow  \mathcal{D}^l(K)U^l = \mathcal{D}^l(K)P^lU^l= P^lU^l = U^l.\] 
Orthonormality of the $\widetilde{\mathcal{D}}^l$ follows directly from the orthonormality of $\mathcal{D}^l$ and  $(U^l)^\dagger U = I$.
Hence, the $\widetilde{\mathcal{D}}^l$  form the natural harmonic basis for diffusion signals living in the quotient space. 
\subsection{Standard Model on \texorpdfstring{$SO(3)$}{SO(3)} and rotational invariants }
To exemplify and motivate the consideration of higher order rotational invariants in the context of $SO(3)$, we here consider generalizations of the Standard Model of diffusion in white matter to triaxial diffusion and B-tensors. Such an extension was recently investigated by Coelho et al in \cite{CoelhoISMRM2026}. As before, $\mathbb{B}$ and $\mathbb{D}$ are the B and D-tensors in their respective principal frames, given by rotations $h$ and $g$ of the lab system. 
The Standard Model of diffusion in white matter\cite{novikovQuantifyingBrainMicrostructure2019} can be formulated in terms of an SO(3) convolution with $f_1(h) = \mathcal{P}(h\hat{e}_3)$, the fiber ODF, and  $f_2(h)= \mathcal{K}\left(\Tr (h\mathbb{B}h^{-1}\,\mathbb{D})\right)$ the kernel (see also \cite{coelhoGeometryCumulantSeries2026}): 
\begin{multline*}
    S(\text{B}) = \int_{\mathbb{S}^2} \mathcal{P}(\hat{n}) \mathcal{K}\left(\Tr (\text{BD})\right)\mathrm{d}\hat{n} \\=\int_{SO(3)}\!\!\mathcal{P}(h\hat{e}_3) \mathcal{K}\left(\Tr (g\mathbb{B}g^{-1}\,h\mathbb{D}h^{-1})\right)\mathrm{d}h\\=\int_{SO(3)} \!\!\mathcal{P}(h\hat{e}_3) \mathcal{K}\left(\Tr (h^{-1}g\mathbb{B}g^{-1}h\,\mathbb{D})\right)\mathrm{d}h \\= \int_{SO(3)} \!\!\mathcal{P}(gh^{-1}\hat{e}_3) 
    \mathcal{K}\left(\Tr (h\mathbb{B}h^{-1}\,\mathbb{D})\right)\mathrm{d}h.
\end{multline*}
We consider generalizations of the Standard Model by allowing triaxial diffusion tensors and encoding tensors -- that is, matrices with 3 distinct eigenvalues. In the former case, the fiber orientation distribution function, fODF, instead becomes a distribution of frames (eigen systems for the local fascicle) $\mathcal{P}(h)$, and the signal for a given b-tensor rotation, $\text{B}=h\mathbb{B}h^{-1}$, becomes

\begin{equation*}    
\label{GSM}
S(g)=\int_{SO(3)}\!\!\mathcal{P}(g h^{-1})\,\mathcal{K}(h)\,\mathrm{d}h
\quad\Longleftrightarrow\quad
S^l=\mathcal{P}^l\,\mathcal{K}^l.
\end{equation*}
This derivation highlights that the lift of the Standard Model to $SO(3)$ follows from allowing general tensor-valued diffusion encoding and an explicit treatment of orientation, independent of whether additional microstructural degrees of freedom--such as triaxial diffusion tensors--are ultimately supported by the data. 

As mentioned above, the dihedral ($D_2$) symmetry of the signal leads to selection rules for $n$ even and $l\neq 1$. 
% Expressed in basis of the reduced space $\operatorname{Im}P^l$, c.f.\ Eq.~(\ref{eq:Stilde}), we see that the convolution law is modified to
% \[
% \widetilde{S}^l = S^lU^l = \mathcal{P}^l\mathcal{K}^lU^l = \widetilde{\mathcal{P}}^l(U^l)^\dagger  \widetilde{\mathcal{K}}^l.
% \]

Furthermore, for axially symmetric $\mathbb{B}$, $\mathcal{K}^l_{mn}\propto \delta_{n0}$, whereas axisymmetric $\mathbb{D}$ leads to $\mathcal{K}^l_{mn}\propto \delta_{m0}$. Combined, this leads to the standard factorization of the Standard Model\cite{novikovQuantifyingBrainMicrostructure2019}. Incidentally, this means that for data acquired with linear b-tensor encoding, the matrix $S^l_{m0}(b)$ for fixed $l$ with $m$ along rows and $b$ along columns,
\begin{equation*}    
\label{eq:LTESM}
S^l_{m0}(b)=\mathcal{P}^l_{mk}\,\mathcal{K}^l_{k0}(b),
\end{equation*}
has rank $> 1$ if the microscopic tensors $\mathbb{D}$ are not axisymmetric -- thereby suggesting a test for the detection of triaxial difusion tensors. Thus, the findings in Ref.~\cite{christiaensNeedBundlespecificMicrostructure2020} indicate that triaxial tensors are at least hard to observe on standard clinical scans.
As for linear b-tensor encoding\cite{novikovRotationallyinvariantMappingScalar2018}, we can define rotation invariant signal scalars as
\begin{equation*}    
\label{eq:rotinv}
S_l^2\equiv \frac1{2l+1}\sum_{mn} |S^l_{mn}|^2 = \frac1{2l+1}\sum_{mn'} |\tilde{S}^l_{mn'}|^2 =\tilde{S}_l^2,
\end{equation*}
where $0\leq S_l\leq 1$. The $S_l$ can be used as new contrasts or for fitting analogously to the RotInv framework\cite{novikovRotationallyinvariantMappingScalar2018}.

\bibliography{SO3powder}%
\end{document}